# Time- and Site-Resolved Kinetic NMR: Real-Time Monitoring of Off-Equilibrium Chemical Dynamics by 2D Spectrotemporal Correlations

Michael J. Jaroszewicz,[1] Mengxiao Liu,[2] Jihyun Kim,[2] Guannan Zhang,[2] Yaewon Kim,[2] Christian Hilty[2,*], Lucio Frydman[1,*]

**Affiliations**

1 Department of Chemical and Biological Physics, Weizmann Institute of Science, 7610001 Rehovot, Israel

2 Chemistry Department, Texas A&M University, 3255 TAMU, College Station, TX, USA

Contributions

C.H. and L.F. conceived the project. M.J.J., C.H. and L.F. developed methodologies, theories and simulations; M.L., J.K., G.Z., Y.K. and C.H. implemented the method. M.J.J., M.L., J.K., Y.K. and G.Z. carried out NMR measurements. All authors discussed the results leading to the final manuscript. M.J.J., M.L., C.H. and L.F. wrote the paper.

Corresponding authors: chilty@tamu.edu, lucio.frydman@weizmann.ac.il

# Abstract

Nuclear magnetic resonance (NMR) spectroscopy provides detailed information pertaining to dynamic processes through line-shape changes, which have been traditionally limited to equilibrium conditions. However, there is a wealth of information to be gained by studying chemical reactions under off-equilibrium conditions –*e.g.*, in states that arise upon mixing reactants that subsequently undergo chemical changes– and in monitoring the formation of reaction products in real time. Herein, we propose and demonstrate a time-resolved kinetic NMR experiment that combines rapid mixing techniques, continuous flow, and single-scan spectroscopic imaging methods, leading in unison to a new 2D spectro-temporal NMR correlation which provides high-quality kinetic information of off-equilibrium dynamics. These kinetic 2D NMR spectra possess a spectral dimension conveying with high resolution the individual chemical sites, correlated with a time-independent, steady-state spatial axis that delivers unique information concerning temporal changes along the chemical reaction coordinate. A comprehensive description of the kinetic and spectroscopic features associated to these spectro-temporal NMR analyses is presented, factoring in the rapid-mixing, the flow and the spectroscopic NMR imaging. An experimental demonstration of this method's novel aspects was carried out using an enzymatically catalyzed reaction, leading to site- and time-resolved kinetic NMR data that are in excellent agreement with control experiments and literature values.

# Introduction

Nuclear Magnetic Resonance (NMR) spectroscopy is capable of studying dynamic processes with atomic-level resolution over a large range of time scales, owing to the extreme sensitivity of the nuclear spin interactions to changes in their surrounding atomic environment. For example fast dynamic processes occurring on $10^{-7}$–$10^{-3}$ s time scales can be studied with suitable relaxation-time measurements,[1] while slower chemical exchanges in the $10^{-3}$–$10^{0}$ s scales can be monitored by analyzing variations in the resulting NMR spectral line shapes.[2,3] Both of these cases are usually studied under equilibrium conditions, whereby dynamic fluctuations happen but chemical species remain under steady-state concentrations. A wealth of information, however, can also be garnered by studying dynamic processes that begin from an off-equilibrium state, and monitoring the chemical dynamics as it proceeds in real time towards equilibrium. So-called stopped-flow NMR techniques have proven particularly effective at studying chemical dynamics in off-equilibrium, without suffering from deleterious spectral artifacts that may otherwise arise from the transient formation and decay of species.[4–7] Stopped-flow NMR has thus been used for studying chemical reactions in polymer and in biomolecular chemistry (*e.g.*, protein and nucleic acid folding);[8] stopped-flow NMR has also been combined with nuclear hyperpolarization methods, which among other benefits allows for the facile detection of unreceptive nuclei.[9] Despite their advantages, stopped-flow NMR methods are of limited use for investigating rapid off-equilibrium chemical reactions, in part due to exchange-lifetime criteria that hinder the observation of individual chemical sites when reactions proceed at rates greater than the frequency-shift differences between the resonances of interest. Further limitations associated with stopped-flow methods involve the mandatory relaxation-delay and dead-time periods needed for spin polarization, mixing, and sample stabilization, as well as the necessity to acquire data from multiple sample batches: all these demands complicate the method, and decrease its capability to analyze the reaction coordinate. There exists, however, another method for monitoring chemical reactions with NMR that can also provide information with respect to the temporal characteristics of chemical reactions processes, which involves the continuous flow of reagents through the NMR probehead.[10–12] These experiments then monitor spectra at usually one or more instants following reagent mixing; additional information with respect to the spatial – and hence the temporal – distribution of reaction processes could arise if the process were monitored as a function of position within the NMR coil. Based on this premise, this work introduces and develops a time-resolved kinetic NMR experiment that combines rapid-mixing and continuous-flow methodologies, with single-scan spectroscopic imaging principles. The result is a novel type of 2D spectro-temporal NMR correlation experiment, capable of providing high-quality, site-resolved kinetic insight about chemical reactions. The principles underlying these novel time- and site-resolved kinetic NMR experiments are discussed in the subsequent sections, including experimental demonstrations, which monitored the depletion and formation of $^1$H methyl resonances evolving from an enzymatically-catalyzed hydrolysis reaction with ms time resolution.

**Results**

Chemical Kinetics from 2D Spectro-temporal NMR Correlations: Overall Principles.

At the heart of the new kinetic- and spectrally-resolved NMR experiment hereby discussed is the co-application of rapid-mixing flow methodologies with ultrafast spatial-spectral imaging schemes. To understand how the combined use of these two modalities allows for the real-time observation of off-equilibrium dynamics, consider a situation in which a chemical reaction is initiated by the rapid mixing of reagents. These reagents then flow with uniform plug-like velocity through a vertical NMR tube with both ends open. The tube is wrapped by an observation (*e.g.* Helmholtz) coil, where flow occurs in the *z* direction parallel to the external field, while the coil is located downstream from the point of mixing (**Figure 1A**). Suppose as well that the spin magnetization of at least one of the reactants is pre-polarized, and that the flow is calibrated in such a manner that regions located near the bottom and top of the NMR coil contain predominantly reactants and products, respectively; *i.e.*, concentration changes and intermediate transient reaction species occur at locations within the NMR coil's field of view. Then, under conditions of stable, continuous, and plug-like flow, each point located at a defined distance away from the point of mixing will contain a unique set of chemical species that correspond to a particular time point along the reaction coordinate (**Figure 1B**). This in turn will give rise to a unique, *z*-dependent NMR spectrum, whereby the unidirectional flow maps in a one-to-one fashion the reaction coordinate of the kinetic process onto the *z* spatial dimension of the NMR coil. Thus, even though the NMR-emitting spins are changing with respect to time due to the kinetics and varying their positions due to the flow, each *z* position will yield a distinct, steady NMR spectrum that corresponds to a definite moment in the reaction process (**Figure 1C**). Given such a scenario, it is possible to retrieve temporally-resolved kinetic information about how the reaction affects individual chemical sites, using a 2D chemical shift imaging approach[13–15] in which gradient-based imaging manipulations resolve, while maintaining chemical site resolution, NMR spectra arising along distinct positions of the $z$ spatial axis. While numerous spectroscopic imaging approaches can do this, speed is also of the essence in this particular case, as the syringe-driven kinetic steady state cannot be maintained indefinitely. Consequently, we decided to employ the echo-planar spectroscopic imaging (EPSI) readout block (**Figure 1D**),[16–18] which delivers 2D spatial/spectral NMR datasets in a single scan, correlating the spins' position along a single-axis gradient (in this case, the direction of the flow along the $z$ axis) in one dimension, with chemical shifts along the orthogonal dimension. The EPSI signal readout occurs while oscillating a bipolar gradient that repeatedly de-phases and re-phases the transverse spin magnetization, resulting in a series of gradient echoes modulated in time by the chemical shift evolution (and by *J*-couplings, here disregarded). Achieving the desired spectrally-resolved 1D images is accomplished by post-processing, which first involves splicing the 1D EPSI signal (FID) into a 2D matrix, whereby each gradient echo is placed in a single row and indexed as a function of a time variable $t_2$ (**Figure 1E**). In this way, the $z$ spatial information is encoded along rows defined by a wavenumber $k_z$, while $t_2$ encodes down each column the chemical shifts – whose evolution is unaffected by the oscillating gradients. After such rearrangement, which may include appropriate alignment of the even and odd gradient echoes, 2D Fourier transformation with respect to $k_z$ and $t_2$ reveals the spectrally-resolved 1D spatial profiles for each chemical site. If the time $T_a$ required to collect each gradient echo (≈100-500 μs) is small with respect to the timescale of the kinetic process, this 2D dataset then reflects the underlying kinetics affecting each chemical site over the course of the reaction (**Figure 1F**).

With these principles as background, the sections that follow delve further into the details of this spectro-temporal 2D correlation approach to monitor chemical kinetics in real time. Proof of the experimental feasibility and validations of this new kinetic NMR methodology are presented in the following Paragraph, which describes the real-time monitoring of the time-dependent depletion and formation of two methyl proton resonances, corresponding respectively to single reactant and product species evolving from an enzymatically-catalyzed, zero-order, off-equilibrium reaction. Then, subsequent Paragraphs within the main and supplementary texts, present a theoretical analysis of the underlying spin dynamics involved in these kinetic spectral imaging experiments. To facilitate a better understanding of the ensuing features of the method, the **Supplementary Information** treats three models that, in increasing order of complexity, show how different facets of the rapid mixing, plug-like flow, and oscillating imaging gradients, end up providing a tool for chemical kinetics. Analytical and numerical derivations are thus compared among themselves and with experiments, and the application of these kinetic NMR methods towards investigating a wider range of rapid off-equilibrium dynamics is discussed. A survey of the current limitations and the challenges affecting these techniques, and potential strategies for how they might be overcome and remedied, are also provided.

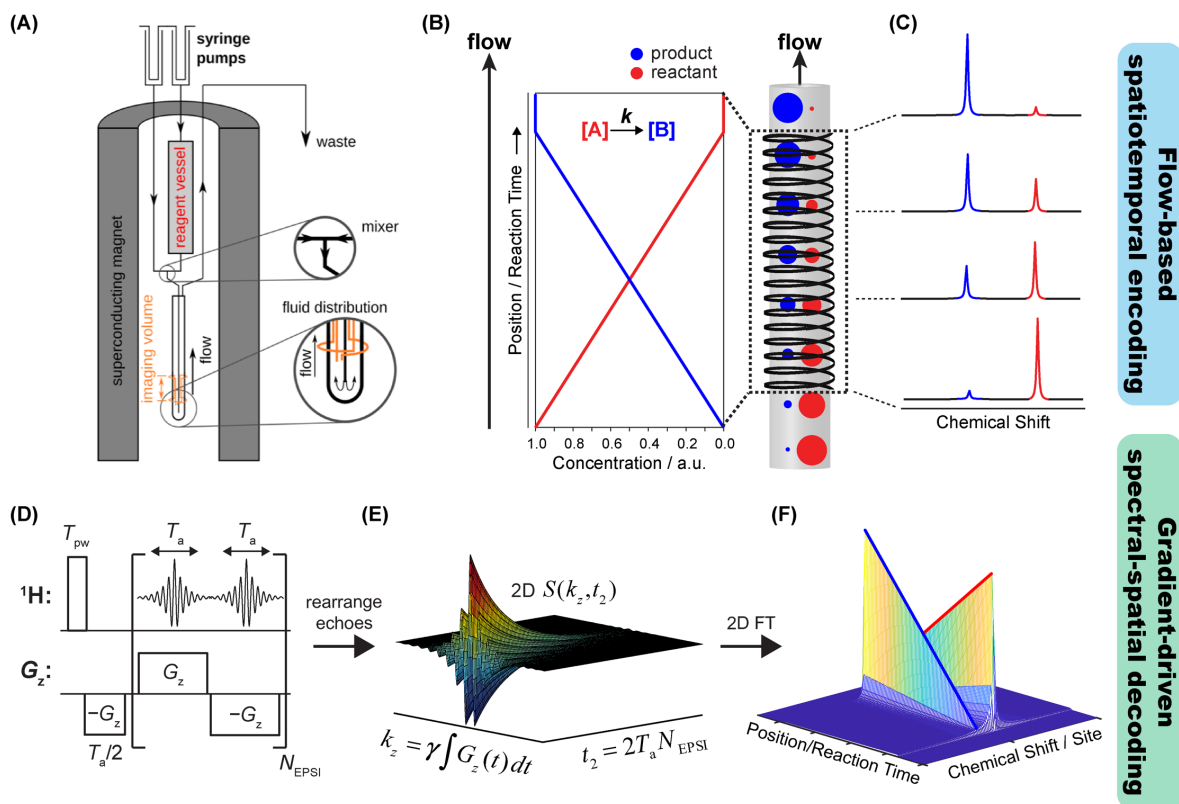

**Figure 1**. **Time-resolved kinetic 2D NMR spectro-temporal correlations arising by combining flow-based spatiotemporal encoding and gradient-driven spectroscopic imaging decoding.** (A) Experimental setup for encoding off-equilibrium kinetics using rapid-mixing and continuous plug-like flow, incorporating a syringe-driven apparatus integrated with a gradient-equipped NMR flow probehead and high-field NMR magnet. (B) Idealized plot of the time-dependent concentrations evolving from a reaction mixture characterized by a zero-order rate law, in which the depletion and formation of the reactant and product are linear functions in time. As this reaction mixture flows through the NMR coil, the plug-like flow encodes each position along the +$z$ direction with a unique, steady-state proportion of reactant and product, as schematically indicated by the size of the red and blue circles, respectively. (C) Each position in space is therefore encoded with a unique NMR spectrum that corresponds to a distinct and definite moment along the reaction process. (D) The echo-planar spectroscopic imaging (EPSI) pulse sequence is used to retrieve and decode the spatially-encoded kinetic information, by resolving NMR spectra as a function of position using an oscillating bipolar gradient-echo train. (E,F) Post-processing of the 1D EPSI time-domain dataset followed by Fourier transformation with respect to the imaging- and time-domain variables $k_z$ and $t$, respectively reveal the spatiotemporally-encoded kinetic NMR spectra.

## 2D Spectro-temporal NMR Correlations: Experimental Tests Using an Enzymatically-Catalyzed Reaction

The enzymatically catalyzed hydrolysis of *N*-α-benzoyl arginine ethyl ester (BAEE) was chosen as a test reaction[9,19] for the method here described (**Figure 2A**). Supporting Information **Figure S.1** and the Methods section provide details on how the continuous-flow apparatus was built and operated. In it, a continuous flow of the substrate (BAEE) was provided through a pre-polarization chamber feeding into the first channel of the mixer, while a solution with the catalyst (the enzyme trypsin) flowed through a second channel without pre-polarization. These flows were driven by at a controllable flow rate, and included a return channel clearing out the products of the reaction outside the NMR probe's field of view. This resulted in the selective polarization of BAEE's $^1$H spins prior to mixing. The mixed solution rapidly flowed through a narrow central tubing to the bottom of an NMR tube. Subsequently, it proceeded to flow upwards with a velocity that was adjusted for the reaction to reach near completion within the sensitive region along the *z*-axis of the NMR coil. This construction allows the device to be top-loaded into a conventional 5 mm NMR probe without requiring special flow-through capabilities. Because the central tubing (0.5 mm inner diameter) was much narrower than the larger return path (42 mm inner diameter), the resulting 70-fold volume difference between the down- and up-flowing solutions allowed us to neglect the former's signal contributions. $^1$H NMR spectra of the flowing reaction mixture were acquired using 2D EPSI and slice-selective imaging pulse sequences, in which the proton methyl resonance belonging to BAEE and the reaction product ethanol (circled in **Figure 2A**) were targeted using selective excitations. See Methods below for additional details on the experimental setup.

**Figures 2B-2D** show experimental 2D $^1$H EPSI data acquired for the flowing reaction, executed with 10 mM BAEE and 7.5 µM trypsin. The depletion of the BAEE reactant peak ($v_{iso}$ = 476 Hz; 1.19 ppm) and build-up of the ethanol product peak ($v_{iso}$ = 437 Hz; 1.09 ppm) are clearly visible in both the 2D spatial-spectral contours (**Figure 2B**), as well as in 1D cross sections taken at the indicated *z* positions from the contour plot (horizontal green lines). Under these experimental conditions ($N_{EPSI}$ = 100, $T_a$ = 0.5 ms, $G_a$ = 20.5 G/cm), the spatial resolution of the EPSI readout

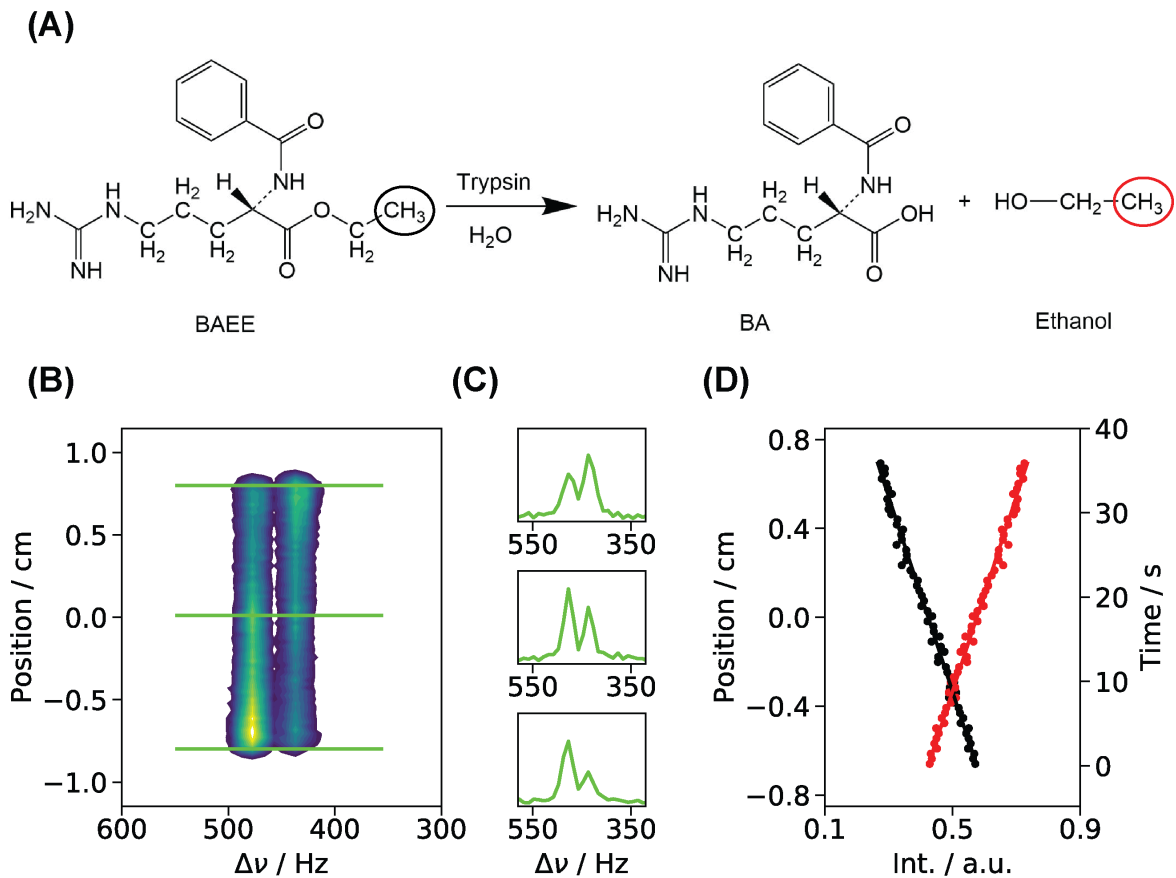

**Figure 2. Kinetic analysis of an enzymatically catalyzed reaction recorded under continuous flow, off-equilibrium conditions *via* 2D spectro-temporal correlations.** (a) Chemical structures of the substrate *N*-α-benzoyl arginine ethyl ester (BAEE), and reaction products benzoyl arginine (BA) and ethanol. The $CH_3$ groups monitored in the experiments are circled in black and red for the reactant and product, respectively. (b) $^1H$ EPSI NMR spectrum of 10 mM BAEE reacting with 7.5 uM trypsin using the pulse sequence in **Figure 1D** ($T_a$ = 0.5 ms, $N_{EPSI}$ = 100, $\Delta z$ = 230 μm, $v$ = 0.33 mL/min, $\Delta t_R$ = 580 ms); (c) green plots show three spectral cross sections taken at the indicated line positions (also green). (d) Integrated spectral intensities (circles) of the reactant (black circles) and product (red circles) resonances plotted against position within the coil. The known rate of flow permits translating positions into reaction times; kinetic fitting of the integrated intensities as discussed in the main text leads to the red and black solid lines, which yielded $k_{cat}$ =11.8 ± 0.5 s$^{-1}$ and $t_0$= 48 ± 3 s for three consecutive acquisitions. A relaxation delay of 5 s and 64 scans were used in these experiments.

is $\Delta z = ca.$ 230 μm, which when factored with the linear flow rate $v$ = 390 μm/s, results in an overall temporal resolution along the reaction coordinate of $\Delta t_R = ca.$ 580 ms. When coupled to the plug-flow assumption and to a knowledge of the flow rate, integration of the spectrally-resolved $^1H$ peaks that EPSI provides in a position-dependent fashion, yields the state of the reaction at different moments in time (**Figure 2D**). To extract the catalysis rate $k_{cat}$ that controls the reaction we realize that, under the conditions assayed, the product formation rate is ultimately proportional to the total enzyme concentration: $\frac{d[P]}{dt} = k_{cat}[E]_0$. Determining $k_{cat}$ is possible by integrating this equation, while realizing that the total amount of reactant and product at any given time, [R] and [P], will equal the total initial concentration $[R]_0$ of BAEE. Assuming that the position-dependent normalized integration values for the signals of the reactant peak $S_R$ (black dots) and product peak $S_P$ (red dots) will be proportional to these concentrations, the position-dependent of the observed resonances can be expressed as:

$$S_R(z) = \frac{1}{[R]_0}([R]_0 - k_{cat} \cdot [E]_0 \cdot (z + z_0)/v) \qquad (1a)$$

$$S_P(z) = \frac{1}{[R]_0}(k_{cat} \cdot [E]_0 \cdot (z + z_0)/v), \qquad (1b)$$

where $[E]_0$, is the initial trypsin concentration, and $z/v$ provides the reaction time $t$. $t_0 = z_0/v$ thus defines the residence time in the "dead volume" occupying the latent space between the point of mixing and the NMR coil, predominantly at the bottom of the NMR tube, and all the other variables have been defined. After three consecutive EPSI NMR measurements, the average $k_{cat}$ and $t_0$ values were determined to be $k_{cat} = 11.8 \pm 0.5$ s$^{-1}$ and $t_0 = 48 \pm 3$ s; given the known flow velocity $v$, the latter corresponds to a dead volume of *ca*. 240 µL. This experimentally measured $k_{cat}$ value is in excellent agreement with previously reported values for this chemical reaction.[9,19]

In addition to these EPSI-based kinetic NMR experiments, a series of slice-selective control experiments were performed (**Figure 3**). In these, several 1D position-dependent (*i.e.*, time-resolved) NMR spectra were collected under similar experimental conditions as described in **Figure 2**, except that the NMR signals (FIDs) were acquired without gradients for longer times (200 ms *vs*. 1000 ms, with the former time given by limitations in the duration over which the gradients could be applied); this lead to a better spectral resolution. A total of 8 consecutive slice-selective 1D $^1$H NMR spectra were recorded at different $z$ positions throughout the sensitive region of the NMR coil (**Figure 3a**), and the corresponding integrated intensities for both the reactant and product peaks (black and red curves, respectively) were calculated and fit in exactly the same manner as in the EPSI measurements (**Figure 3b**). In this case, both $k_{cat}$ and $t_0$ were measured to be 12 s$^{-1}$ and 45 s, respectively; again, in very good agreement with the kinetic parameters obtained with EPSI.

A unique feature of the novel kinetic NMR method introduced in **Figure 1**, is its capability to control the time resolution used to probe the reaction coordinate. For many time-resolved kinetic NMR techniques this time resolution is often dictated by factors outside the control of the experimenter; for spatially-encoded NMR spectra revealed by 2D spectro-temporal correlations, kinetic time resolution is ultimately governed, for a given linear flow rate, by the spatial resolution of the EPSI imaging readout – a tunable and controllable parameter. To demonstrate this added flexibility, **Figure 4** shows how the time resolution can be increased by a factor of *ca*. 6 *vs*. that in **Figure 2**, by increasing both the flow rate by a factor of 3.3× (*v* = *ca*. 1 mL/min), and the EPSI gradient readout spatial resolution by a factor of 2× ($T_a$ = 1 ms and $\Delta z$ = 115 μm). Under these experimental conditions, a reaction-time resolution of *ca*. 86 ms was achieved. Note that this time resolution is shorter than the total time required to collect the entire EPSI gradient echo train, which lasted 200 ms in total (*i.e.*, $N_{EPSI}$ = 100 for $T_a$ = 1 ms). This added flexibility for monitoring kinetic processes with high temporal resolution comes at a price, however, which is evidenced by the *ca*. 3× reduction in signal-to-noise (SNR) for **Figure 4** *vs*. **Figure 2**. This results from the combined effects of the increased flow rate, and of the higher spatial resolution of the gradient readout. Despite these SNR reductions, it is still possible to achieve similar kinetic fits as those ascertained in **Figure 2**: three independent measurements here yielded $k_{cat}$ = 11.7 ± 0.3 s$^{-1}$ and $t_0$ = 12 ± 1 s. As before, 1D slice-selective pulse-acquire measurements carried out under the same experimental conditions yield virtually identical kinetic parameters (**Figure 4d-4e.**). **Supplementary Note 2** and supporting **Figures S.4-S.6** contained therein provide a more in-depth analysis of the combined effects of the imaging spatial resolution and flow rate on the overall SNR.

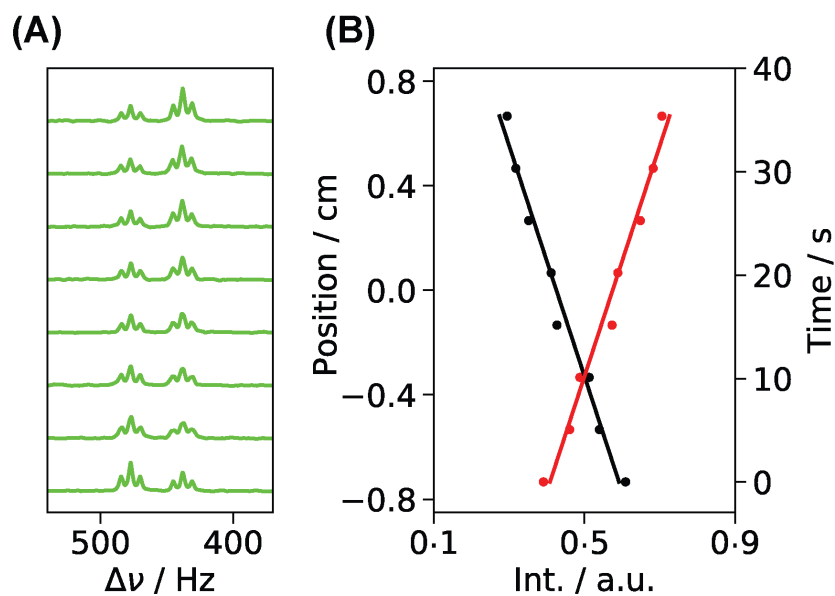

**Figure 3**. **1D Slice-selective NMR spectra of the enzymatically-catalyzed hydrolysis reaction described in Figure 2.** (A) Slice-selective (2 mm) $^1$H NMR spectra of the methyl region acquired at 8 distinct positions along the NMR coil, collected under similar experimental conditions as those in **Figure 2,** but utilizing a 1 second acquisition time. The improved resolution reveals the *J*-coupling of the sites. (B) Integrated signal intensities for the reactant (black circles) and product (red circles) plotted as a function of both position and reaction time against the lines of best fit from the kinetic analysis ($k_{cat}$ =12.0 s$^{-1}$ and $t_0$= 45 s).

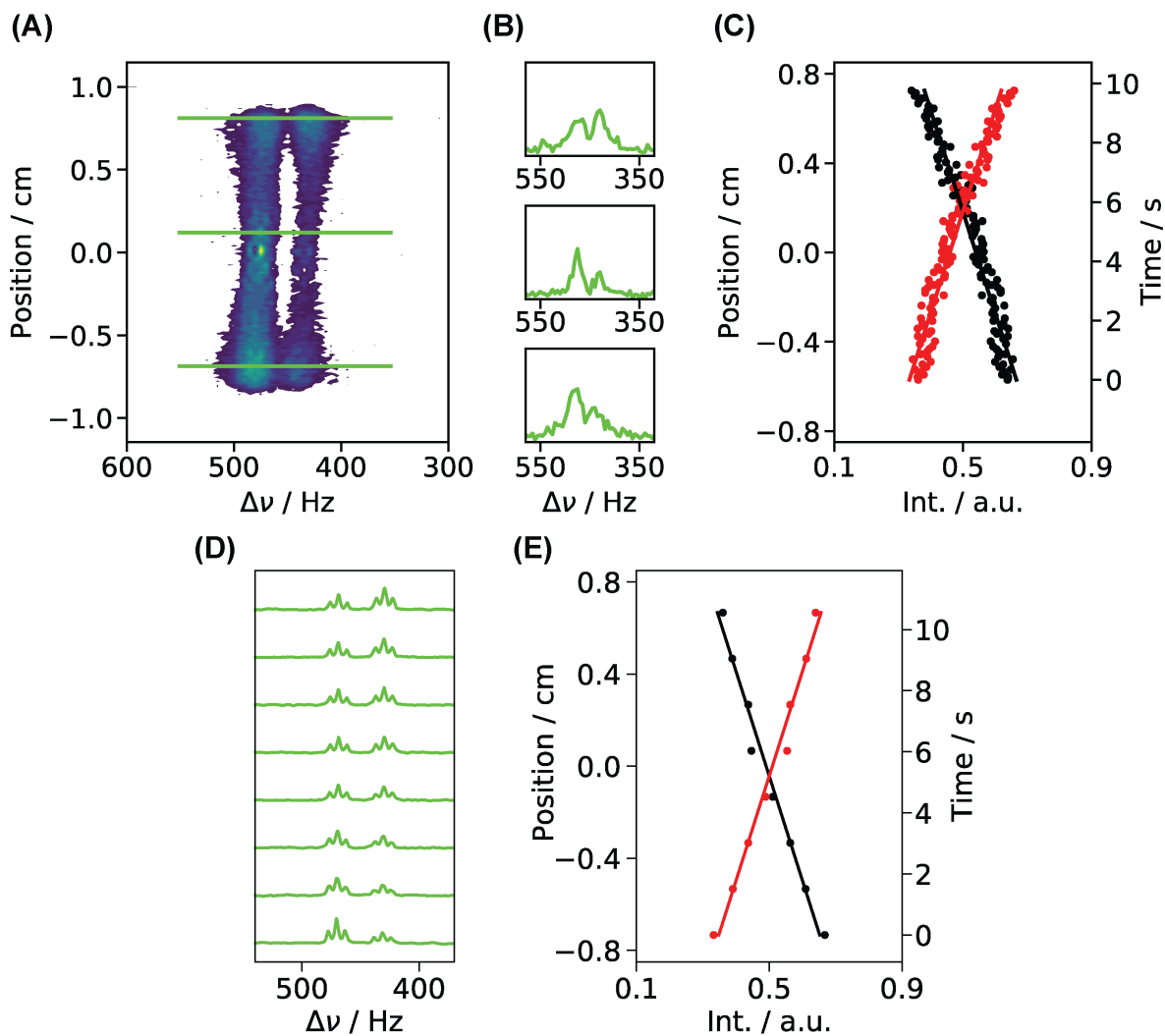

**Figure 4. High-resolution ($\Delta t_R$ = 86 ms) time-resolved kinetic NMR spectra of an enzymatically-catalyzed hydrolysis reaction in off-equilibrium.** (a) $^1$H EPSI NMR spectrum and (b) associated spectral cross sections acquired with a reaction-time resolution of $\Delta t_R$ = 86 ms under conditions of continuous flow for a reaction mixture of 25 μM trypsin and 10 mM BAEE. The following experimental parameters were used in the collection of (a): $N_{EPSI}$ = 100, $T_a$ = 1 ms, $G_z$ = 20.5 G/cm, TD = 80000 (200 complex pts/echo), and $v$ = 1200 μm/s (1 mL/min). 240 scans were collected with D1 = 5 s. (c) Integrated signal intensities for the reactant (black circles) and product (red circles) plotted as functions of both position and reaction time against the lines of best fit from the kinetic analysis (solid lines): $k_{cat}$ =11.7 s$^{-1}$ and $t_0$ = 11 s. (d) $^1$H slice-selective pulse-acquire NMR spectra acquired at 8 distinct positions across the NMR coil and under similar experimental conditions as those in (a)-(c): 25 μM trypsin, 10 mM BAEE, and the flow rate $v$ = 1 mL/min. The kinetic fitting results obtained from these data are: $k_{cat}$ = 11.7 s$^{-1}$ and $t_0$ = 12 s.

## The Spatio-temporal Encoding and Spectral-spatial Decoding Processes

The information of 2D spectro-temporal kinetic correlations arises from the simultaneous action of flow-based and gradient-based encoding and decoding schemes. As such they exhibit, in addition to chemical kinetic information, dependencies on both the flow and imaging characteristics of the experiment. An analysis of these characteristics is thus justified; the present paragraph examines how the chemical kinetics interacts with the flow and with the specific EPSI experiment, to govern the spin evolution eventually leading to 2D spectro-temporal NMR correlation spectra. We consider a continuous plug-like flow of an off-equilibrium reaction mixture that occurs before and during the acquisition of an FID, which can be excited by a single resonant radio-frequency (RF) pulse or by an EPSI readout. It is also assumed that the chemical reaction is described by an irreversible zero-order rate law ($A \xrightarrow{k} B$), that is instantaneously initiated in a reaction vessel located immediately upstream from the NMR coil; *i.e.*, for simplicity we assume there is no latent volume between the point at which the reaction is initiated and the entrance into the NMR coil. At the coil thus enters only the reactant **A**, with the product **B** forming progressively down the length of the coil. A linear flow at rate *v* gives a distribution of **A** and **B** longitudinal magnetization that is a function of both position and time, which exists in a dynamic equilibrium for all fixed positions *z* within the NMR coil when $vt > L_z$, where *t* is the elapsed time post-mixing, and $L_z$ is the length of the NMR coil. Describing the spin dynamics of **A** and **B** under these combined conditions of plug-like flow and zero-order chemical kinetics is easier if first the dynamics of **A** and **B** are determined in the absence of flow. A description for such a case is presented in **Supplementary Note 3**. Denoting $S_A(t)$ and $S_B(t)$ the signals then emitted by **A** and **B** (still in the absence of flow or field gradients) after a hard-pulse excitation of the entire sample at a time *t* = 0 coinciding with the time at which the reaction is initiated, this **Supplement** shows that

$$S_A(t) \propto \begin{cases} (A_0 - kt) \times \exp(\lambda_A t), & t < t_{max} \\ 0, & t \geq t_{max} \end{cases} \quad (2a)$$

and

$$S_B(t) \propto \begin{cases} k \dfrac{\exp(\lambda_A t) - \exp(\lambda_B t)}{\lambda_A - \lambda_B}, & t < t_{max} \\ k \dfrac{\exp(\lambda_A t_{max}) - \exp(\lambda_B t_{max})}{\lambda_A - \lambda_B} \times \exp(\lambda_B \cdot (t - t_{max})), & t \geq t_{max} \end{cases} \quad (1b)$$

These signals are proportional within identical factors to the time-dependent complex transverse magnetizations of **A** and **B** post-excitation, and they depend on the post-excitation time *t*, **A**'s initial concentration $A_0$ (in M), and the zero-order reaction rate constant $k$ (in M s$^{-1}$). Here $\lambda_{A/B} = i\omega_{A/B} - R_{2A/B}$, where $\omega_{A/B}$ and $R_{2A/B}$ are the **A/B**-spin chemical shift (in rad/s) and transverse relaxation rates (in s$^{-1}$) respectively, and $t_{max}$ is a post-excitation time at which the chemical transformation is completed and all of **A** has been depleted. **Supplementary Note 3** provides a derivation and detailed analysis of these equations, to which the presence of additional **B** at time *t* = 0 can also be added without excessive complications.

In order to add the effects of continuous flow onto Equations (2) and account for positional changes in the magnetizations' distributions, a spatial translation operator $\mathcal{T}(z)$ representing the action of the flow, was included. The resulting post-excitation, flow-dependent, kinetic NMR signals for **A** and **B** – still in the absence of pulsed magnetic field gradients – are then given by integrals over all $0 \leq z \leq L_z$ positions within the active coil volume, as:

$$S_A(t) \propto \begin{cases} \int S_A(z,t)dz \propto \int M_+^A(z,t) \times \exp(\lambda_A t)\, dz, & t < \dfrac{A_0(z)}{k} \\ 0, & t \geq \dfrac{A_0(z)}{k} \end{cases} \quad (3a)$$

$$S_B(t) \propto \begin{cases} \int M_+^B(z,t) \times \exp(\lambda_B t)\, dz + \int \mathcal{T}\left\{\dfrac{k}{\lambda_A - \lambda_B}\right\} \times (\exp(\lambda_A t) - \exp(\lambda_B t))\, dz, & t < \dfrac{A_0(z)}{k} \\ \int M_+^B(z,t) \times \exp(\lambda_B t)\, dz + \int \mathcal{T}\left\{k\dfrac{\exp(\lambda_A t_{max}(z)) - \exp(\lambda_B t_{max}(z))}{\lambda_A - \lambda_B}\right\} \times \exp(\lambda_B(t - t_{max}(z)))\, dz, & t \geq \dfrac{A_0(z)}{k} \end{cases} \quad (3b)$$

$M_+^A(z,t)$ and $M_+^B(z,t)$ represent respectively the **A** and **B** transverse magnetization amplitudes, which now have distinct kinetic and flow-based dependencies that take the form:

$$M_+^A(z,t) \propto \begin{cases} \mathcal{T}(A_0(z)) - kt = \mathcal{T}\left\{A_0 - \dfrac{k}{v}z\right\} - kt, & t < \dfrac{A_0(z)}{k} \text{ and } t < \dfrac{L_z - z}{v} \text{ and } t < \dfrac{z}{v} \\ 0, & t \geq \dfrac{A_0(z)}{k} \text{ or } t \geq \dfrac{L_z - z}{v} \text{ or } t \geq \dfrac{z}{v} \end{cases} \quad (4a)$$

and,

$$M_+^B(z,t) \propto \begin{cases} \mathcal{T}(B_0(z)) = \mathcal{T}\left\{\dfrac{k}{v}z\right\}, & t < \dfrac{A_0(z)}{k} \text{ or } t \geq \dfrac{A_0(z)}{k} \text{ and } t < \dfrac{L_z - z}{v} \text{ and } t < \dfrac{z}{v} \\ 0, & t \geq \dfrac{L_z - z}{v} \text{ or } t \geq \dfrac{z}{v} \end{cases} \quad (4b)$$

Here $\mathcal{T}$ is the spatial translation operator taking as input $A_0(z)$ and $B_0(z)$, where $A_0(z)$ and $B_0(z)$ represent the spatially-dependent dynamic equilibrium magnetizations of **A** and **B**, respectively, at $t = 0$ immediately preceding excitation. $\mathcal{T}$ then translates these magnetizations from their initial positions $z$ at $t = 0$, by an amount $\delta z = vt$ as a function of the FID acquisition time. All other symbols possess similar definitions as given earlier. Notice that these Equations rely on $v$ as a unique, constant plug-flow rate, in order to relate the evolution time $t$ for a spin with its changing position within the NMR coil. Notice as well that in addition to the boundary condition accounting for the kinetic consumption of – a now spatially-dependent – $A_0(z)$ by the chemical reaction proceeding at a rate $k$, (*i.e.*, by making the $t_{max}$ of Equations (2) equal to $t_{max}(z) = \frac{A_0(z)}{k}$), a new set of flow-dependent boundary conditions are introduced in Equations (4). These account for both excited magnetizations flowing out of the coil, which zeroes-out contributions for certain spins at times $t \geq \frac{L_z - z}{v}$ as they have exited the coil's field of view, as well as for the in-flow of non-excited (even if pre-polarized) magnetizations flowing into the coil at times $t \geq \frac{z}{v}$. **Figure 5** clarifies further the time- and position-dependent magnetizations of **A** and **B** subjected to the flow- and kinetic-dependent boundary conditions given by Equations (4), for $k = 0.375$ M s$^{-1}$ and a linear flow rate of $v = 0.53$ cm/s. **Figures 5A and 5B** show the time- and space-dependent magnetizations within a schematized NMR coil, illustrating the depletion and growth of the reactant (red) and product (blue) as a function of $z$.

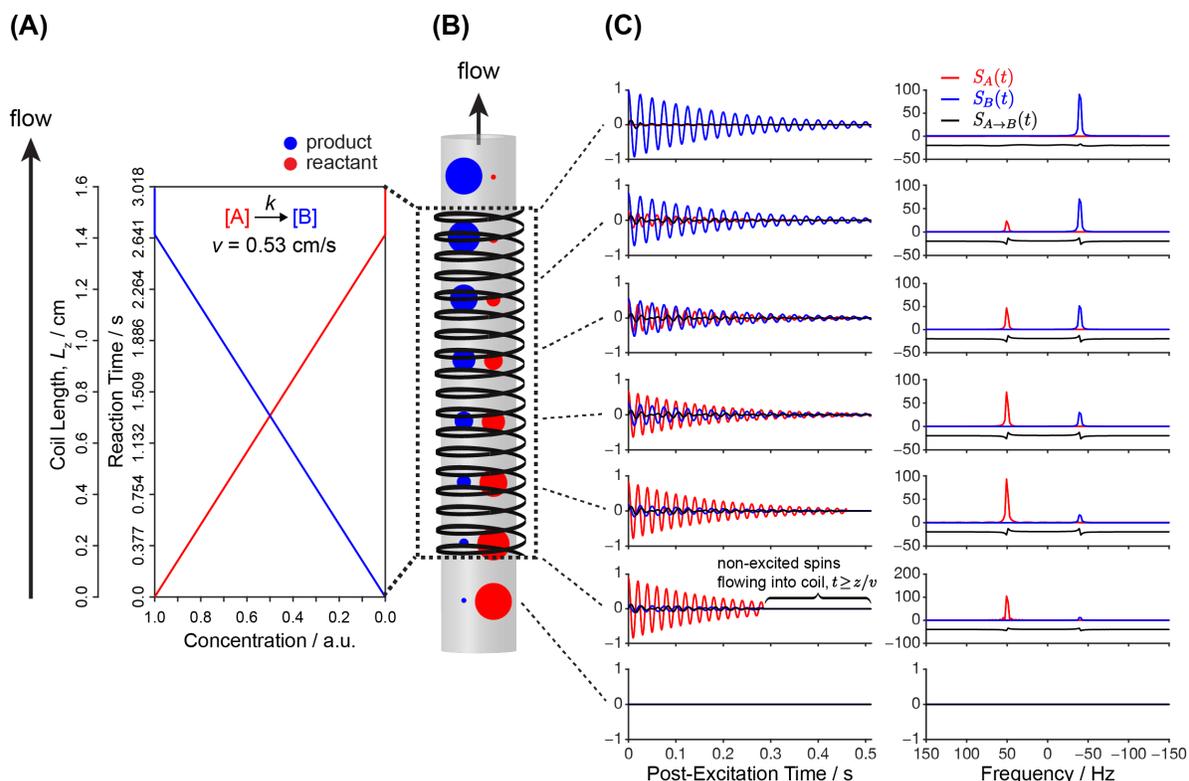

**Figure 5**. **Flow-based spatiotemporal encoding of chemical kinetics.** (A) Zero-order kinetic profile ($k$ = 0.375 M s$^{-1}$) assumed for an off-equilibrium reaction (**A**→**B**), plotted as a function of both reaction time and position for $v$ = 0.53 cm/s and $L_z$ = 1.6 cm discretized over 256 unique $z$ positions. (B) Schematic representation of the spatiotemporally-encoded chemical kinetics along the $z$ spatial coordinate of the NMR coil, that results from such plug-like flow. The relative spatially-dependent concentrations of **A** and **B** are schematized with the red and blue circles, respectively, with their corresponding simulated time-resolved 1D NMR FIDs and spectra shown in (C), which were generated using Equations (3) of the main text.

As discussed in the **Overall Principles** section, the unique proportion of **A** and **B** for every position $z$ will correspond to definite moments along the reaction coordinate, reflecting in turn in the NMR correlations observed by 2D EPSI. The idealized (*e.g.*, turbulence- and diffusion-free) time- and frequency-domain responses of these sites, are in **Figure 5C**. Shown by red and blue traces are the NMR responses of **A** and **B**, respectively, that originate solely from the equilibrium magnetization established pre-excitation; shown in black are the NMR spectra of **B** signals that were generated post-excitation *via* the zero-order kinetics. Notice that the latter contain contributions from both sites, and a mixed-phase line shape reflective of the multiple $t' > 0$ instants at which the **A**➔**B** interconversion took place. These simulations also demonstrate that the contributions of these signals for the reaction conditions designed in the experiments above, are much smaller than the contributions present at $t = 0$ pre-excitation, and hence they are not detectable; they would, however, be noticeable for faster reaction rate constants $k$. Note as well that the direction of the plug-like flow, assumed to proceed in a vertical, +$z$ direction, causes a premature FID zeroing for both $M_+^A$ and $M_+^B$ located at $z$ positions nearest the bottom of the NMR coil at post-excitation times $t \geq \frac{z}{v}$ (lower traces in the figure). This manifests in the NMR spectra as a line broadening.

While these analytical simulations can be used to predict the kinetically-resolved 1D NMR spectra for distinct positions, a conventional 1D pulse-acquire experiment cannot deconvolve the individual spatially-dependent contributions to the **A** and **B** signals. Only spatially-selective acquisitions employing field gradients, can decode and readout the kinetic NMR information encoded in this fashion. The details of the latter are described in the next Paragraph.

## Real Time Gradient-Driven Decoding of the Spatially Encoded Chemical Kinetics

The spectro-temporal NMR method in **Figure 1** relies on EPSI to retrieve the aforementioned spatially-encoded kinetic information, with chemical shift resolution. Modelling the effects of the EPSI time-dependent oscillating gradient readout with the zero-order chemical kinetics and the continuous plug-like flow can be accomplished by appending to the treatment in the preceding Paragraph, the relevant gradient-induced shifts that affect each spin as it changes position due to the flow, and as it changes chemical environment due to the chemical kinetics. In addition, the various boundary conditions detailed in the previous Paragraph need to be respected. Once again the NMR signal of the starting reactant **A** as a function of all these factors is easier to summarize, and is given by

$$S_A(t) \propto \begin{cases} \iint M_+^A(z,t) \times \exp(\lambda_A t) \times \exp\left(i\gamma \int_0^t G(t') \cdot z(t')dt'\right) dz, & t < \dfrac{A_0(z)}{k} \\ 0, & t \geq \dfrac{A_0(z)}{k}, \end{cases} \quad (5a)$$

where $M_+^A(z,t)$ is given by Equation (4a), $\gamma$ is the gyromagnetic ratio of the imaged spins in rad·Hz/T, $G(t')$ is the time-dependent gradient amplitude in G/cm, and $z(t') = z + vt'$ is the time-dependent position of a given flowing spin at time $t'$ that had an initial position $z$, with $0 \leq t' \leq t$. The new exponent in Equation (5a) highlights the fact that spins are undergoing motion while being imaged – a topic, which has been extensively discussed in the NMR and MRI literature.[15,20–23] As is well known from such studies, this gradient/flow combination can result in signal attenuations and/or in secular-like shifts of the signals. **Supplementary Note 4** investigates this phenomenon in greater detail as it pertains to spectrotemporally-encoded NMR correlations, and demonstrates that in a pre-phased EPSI readout of the kind given in Equations (5), the spin evolution is compensated against the effects of plug-like flow. Therefore, the resulting NMR signals are devoid of flow-induced frequency shifts (even if they will be affected by flow-dependent broadening related to out-flowing of polarized spins and in-flowing of unpolarized ones, *vide supra*).

The corresponding expression for **B** is composed in an analogous fashion as was carried out for **A**, in which each unique chemical shift term in Equation (S1.2) and Equation (S1.3) (**Supplementary Note 3**) is multiplied by its corresponding gradient-induced shift. After some algebra (see S**upplementary Note 5** for the full derivation), the following piece-wise expression is obtained:

$$S_B(t) \propto \int_Z M_+^B(z,t) \times \exp(\lambda_B t) \times \exp\left(i\gamma \int_0^t G(t') \cdot z(t') dt'\right) dz$$

$$+ \int_Z k \frac{\exp(\lambda_A t) - \exp(\lambda_B t)}{\lambda_A - \lambda_B} \times \exp\left(i\gamma \int_0^t G(t') \cdot (z + vt') dt'\right) dz, \quad t < \frac{A_0(z)}{k} \quad (5b)$$

$$S_B(t) \propto \int_Z M_+^B(z,t) \times \exp(\lambda_B t) \times \exp\left(i\gamma \int_0^t G(t') \cdot z(t') dt'\right) dz$$

$$+ \int_Z k \frac{\exp(\lambda_A t_{max}) - \exp(\lambda_B t_{max})}{\lambda_A - \lambda_B} \times \exp(\lambda_B \cdot (t - t_{max})) \quad (5c)$$

$$\times \exp\left(i\gamma \int_0^t G(t') \cdot (z + vt') dt'\right) dz, \quad t \geq \frac{A_0(z)}{k}$$

where $M_+^B(z, t)$ is defined in Equation (4b), and the spatial integral extends over the relevant field-of-view. Equations (5) therefore describe all of the relevant time-dependent spin- and space-based dynamics relevant to 2D spectrotemporal NMR kinetic correlations, and permit analytical simulations of the aforementioned experimental datasets. **Supplementary Note 5** provides a more thorough analysis of the resulting theoretical EPSI line shapes described with Equations (5), by exploring three limiting cases in which: i) the spins evolve in the absence of the chemical kinetics (*i.e.*, $k = 0$) under the effects of both field gradients and flow; ii) stopped-flow conditions are employed during the EPSI readout (*i.e.*, $v \neq 0$ pre-excitation and $v = 0$ post-excitation); iii) $v > k$ leading to measurable flow-induced broadening under conditions of continuous flow during FID acquisition.

## Discussion

A novel, versatile kinetic method for monitoring off-equilibrium chemical transformations in real time by NMR, was proposed and demonstrated. The method relies on the synergistic and combined application of rapid-mixing, continuous flow, and single-scan spatial-spectral acquisition schemes. It retrieves site-resolved information that would be hard or impossible to obtain by other means. To test these ideas a custom syringe-driven flow apparatus allowing for the rapid mixing of selectively pre-polarized reagents was built, leading to a stable, continuous flow of reaction mixtures through a standard 5 mm commercial NMR probehead. The continuous plug-like flow served to map in a one-to-one fashion the time coordinate of the off-equilibrium reaction onto the *z* spatial coordinate of the NMR coil. The flow-derived spatially-encoded kinetic NMR information was then retrieved by using echo-planar spectroscopic imaging in which the site- and time-resolved resonances for the reactants, products, and transient species from every *z* position in space were measured in a single scan. Experimental validation of this technique was performed by monitoring the enzymatically-catalyzed hydrolysis of an ester and the corresponding formation of ethanol, in which the methyl proton resonances of both were selectively targeted. The custom-built flow apparatus and associated hardware gave precise control over the rate and the nature of the fluid flow profile. It allowed for the collection of quality slice-selective and EPSI NMR data averaged over minutes, and executed under different flow rates and spatial encoding conditions. The encoded kinetic NMR information could be extracted from these data, and agreed remarkably well with both reported literature values and control experiments. This kind of experiment required no modifications on widely available NMR hardware, relying on a mixing device designed to be loaded instead of an NMR tube into a conventional 5mm probe.

In principle, this combined approach based on using flow to spatially encode chemical transformations followed by a joint NMR/MRI decoding of the spectral and spatial information, can deliver information about reaction coordinates with intriguingly high kinetic resolution. From an experimental standpoint, its most important limitations appear to be dictated by sensitivity considerations, and by the stability of the flow. The former arises as probing faster kinetic timescales will be associated with resolving smaller spatial elements, leading to decreased signals; the latter due to the fact that although the system is assumed to be spatially at a steady state, deviations in plug flow and/or turbulences will blur the kinetic information. In addition to these experimental constraints, this study explored the theoretical limits of this new method, using a novel analytical model complemented by numerical simulations. These described the NMR spin dynamics for an off-equilibrium zero-order chemical reaction, while accounting for the EPSI acquisition under the combined effects of continuous flow and chemical kinetics. This model showed that joint MRI/NMR acquisitions of flowing off-equilibrium reactions contain signal contributions from two distinct sources: one includes the reactants and products established pre-excitation, and the other arising from the products created over the course of FID acquisition post-excitation. For the experimental examples presented here, the former is the dominant contribution to the NMR line shape. The latter's contributions, however, will become relevant when faster reaction rates are examined, as a greater proportion of product gets created over the course of FID acquisition itself. The mixed phase character of these resulting line shapes encodes rich information pertaining to the chemical kinetics, that can reveal insight unavailable by existing approaches. Our theoretical model also showcased the conditions by which flow-based features arising from the inflow of unexcited, NMR-silent spins and the effluing of polarized sources, will affect the resulting line shapes. These will manifest as spectral broadenings, whose removal will require special consideration pertaining to the experimental setup when investigating kinetics governed by faster reaction rates. This may include exploiting reduced field-of-view imaging, in combination with RF pulses having increased spatial excitation bandwidths. On-going work includes the targeting of faster reaction kinetics through the combined use of increasingly high-resolution spatial readouts, faster linear flow rates, and advanced modeling of the kinetics. A crucial requirement for achieving higher temporal reaction resolutions will be adequate sensitivity, which we are striving to achieve by incorporating hyperpolarization. Alternatively, the reliance on microfluidic components might provide the sensitivity/sample volume compromise needed for a more widespread application of this new method.

## Methods

### The Mixer Device

A mixer device was designed and constructed for continuous flow NMR with the capability to pre-polarize the nuclear spins of one of the reactants. The device was constructed to fit within a narrow-bore NMR magnet equipped with a commercial shim system. The mixer contains two inlet channels for the two reactants (**Figure S.1**). One reactant passes through the middle inlet leading to the pre-polarization chamber (2.7 mL volume), which provides an increased residence time of nominally 3 s at a flow rate of 50 mL/min in order to polarize the spins. The second reactant is injected through the other inlet and mixes with the first reactant downstream of the pre-polarization chamber. Immediately below, the mixture flows through a leading tube (with 1/32 and 1/50 in. outer and inner diameters, respectively, IDEX Health & Science, Oak Harbor, WA) to the bottom of a standard 5 mm NMR tube inserted into the NMR probe. The reaction mixture then flows upward in the direction parallel to the main magnetic field through the NMR coil. The spent reaction solution eventually flows out from the return channel of the mixer and is collected in a waste container. The device was made from a cylinder of Delrin plastic with an outer diameter of 25 mm and a length of 75 mm (Online Metals, Seattle, WA). Channels of 1.6 mm width and depth were machined by computer numerical control (CNC) into a flat surface milled into the cylinder (letter A in **Figure S.1**). The channels were sealed with a single gasket sheet made from nitrile rubber (McMaster-Carr, Elmhurst, IL). The cover (B) was attached with brass screws (4-40 unified thread standard). The NMR tube was attached below the mixer, using a nitrile O-ring between pieces (A) and (C), and a retaining ring permanently glued to the NMR tube between pieces (C) and (D). Mixing was improved with a cotton string inside a section of the leading tube (green tubing, 1/16 inch outer and 1/32 in. inner diameter, **Figure 1b**).

Pulse Sequences and Data Processing

All NMR spectra were recorded using a triple resonance TXI probe installed in a 9.4 T NMR spectrometer (Bruker Biospin, Billerica, MA). The EPSI pulse sequence[13] (**Figure 1d**) used an excitation pulse that was frequency-selective to excite 400 Hz of the $^1$H methyl region (using a Pc9_4_90 shape) in $\tau_{pw}$ = 18.78 ms, which was experimentally optimized to give both uniform excitation in the frequency domain as well as in the spatial domain (**Figure S.2**). An EPSI pre-phasing period of $T_a/2$ = 0.25 or 0.5 ms was used along with associated readout times lasting $T_a$ = 0.5 or 1 ms, with all gradients having an amplitude of $|G_z|$ = 20.5 G/cm. The pair of bipolar gradients were looped $N_{EPSI}$ times, in which $2N_{EPSI}$ gradient echoes were recorded. All EPSI NMR spectra were generated by rearranging the 1D gradient echo trains into 2D matrices, in which only the odd echoes were used for processing. Processing included Gaussian multiplication and zero-filling in both dimensions, followed by 2D Fourier transformation with respect to $k_z$ and $2N_{EPSI}T_a$ along each row and column, respectively. All data are presented in magnitude mode. A spatial resolution of $\Delta z = 1/\gamma G_z T_a$ = 115 μm was used in the EPSI measurements unless stated otherwise. The spectral width is given by twice the spacings between adjacent gradient echoes (*i.e.*, DW = 1 or 2 ms and SW = 1000 or 500 Hz, respectively), which gives a spectral resolution of 10 or 5 Hz/pt. A G4 Gauss-Cascade pulse[14] with a bandwidth of 15 kHz was used for slice-selective excitation experiments in the presence of a *z* gradient with amplitude $|G_z|$ = 17.62 G/cm, resulting in a slice thickness of *ca*. 2 mm. A total of 8 slices within the sensitive region of the NMR coil were measured successively using frequency offsets covering a total of 105 kHz with a step size of 15 kHz. A water suppression sequence consisting of an EBURP 2.1 shaped π/2 pulse followed by randomized pulsed field gradients $G_x$, $G_y$ and $G_z$, were applied before the excitation pulse. Data acquisition lasted 1 s with a total of 6400 complex points acquired in a single slice. A simple spectral normalization procedure was carried out for all EPSI NMR datasets. This procedure corrected for discrepancies between $^1$H peak intensities in the EPSI $^1$H NMR spectra and conventional 1D $^1$H NMR spectra that were collected on a stationary (*i.e.*, non-flowing) mixture of BAEE and ethanol. In this case, the methyl peak corresponding to the ethanol had a larger intensity than that of BAEE in the 1D $^1$H NMR spectrum, whereas the reverse was the case in the EPSI NMR spectrum. Since the total concentration of the reactant and the product remains constant for a given position under conditions of continuous flow, the peak integrals of both were normalized before kinetic fitting according to: $\hat{S}_R = S_R/(S_R + S_P)$ and $\hat{S}_P = S_P/(S_R + S_P)$, whereby the R and P subscripts refer to the reactant and product integrals respectively. Furthermore, due to the characteristic 1D spatial coil profile recorded in EPSI NMR spectra whereby the signal intensity falls off at the edges, only signals originating from within the ± 0.7 cm region from the center of the coil were considered for kinetic analysis (**Figure S.3**).

Sample Preparation

The substrate solution was prepared by dissolving *N*-α-benzoyl arginine ethyl ester (BAEE, TCI Chemicals, Tokyo, Japan) in 50 mM pH 7.6 phosphate buffer. The enzyme trypsin (AMERSCO, Road Solon, OH) was dissolved in the same buffer at a concentration of 7.5 or 25 µM (depending on the flow rate). 10% $D_2O$ was added to the substrate solution for frequency locking. The substrate and enzyme solutions were injected into the mixer device in a 1:1 volume ratio with the use of a syringe pump (Fusion 200 Touch, Chemyx, Stafford, TX).

## Data Availability
The data sets generated and analyzed during the current study are available from the corresponding authors on request.

## Code Availability
The MATLAB-based code used for simulating experimental datasets is available from the corresponding authors on request.

## Acknowledgements

LF holds the Bertha and Isadore Gudelsky Professorial Chair and Heads the Clore Institute for High-Field Magnetic Resonance Imaging and Spectroscopy, whose support is acknowledged. This work was supported by the US-Israel Binational Science Foundation program (Grant 2014316), Israel Science Foundation Grant 965/18, the Perlman Family Foundation and the Welch Foundation (Grant A-1658).

## Ethics declarations

The authors have no competing interests to declare.

## Supplementary Information

Supplementary Information: Additional information on the experimental setup; analytical derivations for the effects of kinetics, flow and oscillating gradients on the signals arising from an NMR experiment; numerical simulations of the various scenarios.

Supporting information for

# Time- and Site-Resolved Kinetic NMR: Real-Time Monitoring of Off-Equilibrium Chemical Dynamics by 2D Spectrotemporal Correlations

Michael J. Jaroszewicz,[1] Mengxiao Liu,[2] Jihyun Kim,[2] Guannan Zhang,[2] Yaewon Kim,[2] Christian Hilty[2,*], Lucio Frydman[1,*]

**Affiliations**

1 Department of Chemical and Biological Physics, Weizmann Institute of Science, 7610001 Rehovot, Israel

2 Chemistry Department, Texas A&M University, 3255 TAMU, College Station, TX, USA

Supplementary Note 1. **Additional experimental details concerning the mixer device, the methyl-selective excitation scheme, and the kinetic analysis.**

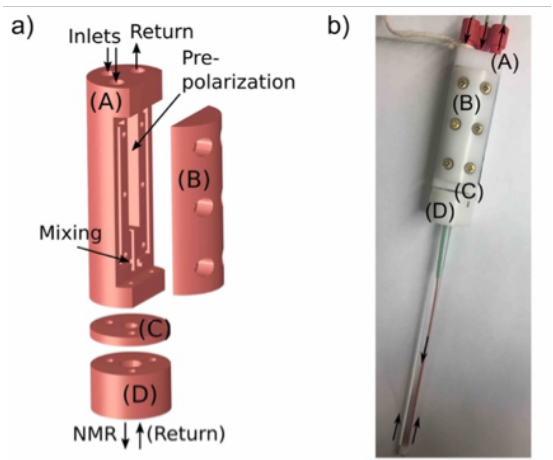

**Figure S.1**. **Schematics of the custom-built mixer device.** (a) Expanded view of the device for rapid mixing flow-NMR spectroscopy showing the channel substrate (A), cover plate (B), as well as the NMR tube adapter consisting of the sealing plate (C) and the tube retainer (D). (b) The fabricated device integrated with a standard 5 mm NMR tube and fluid connectors.

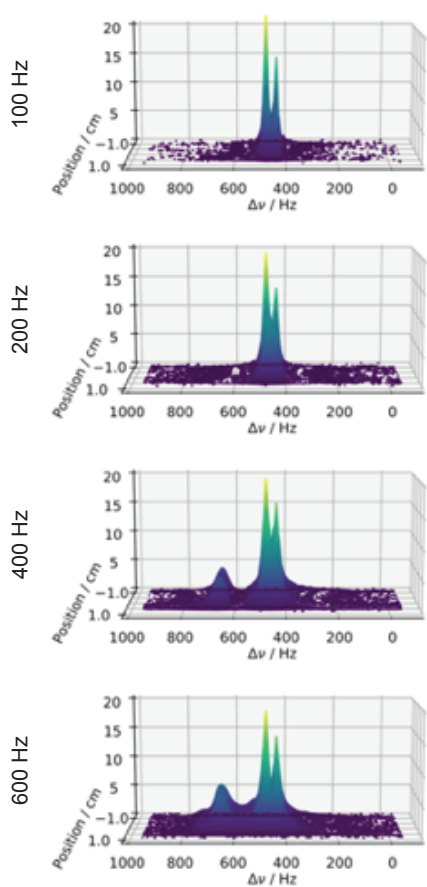

**Figure S.2**. **Experimental optimization of the $^1$H methyl selective excitation pulse.** Experimental $^1$H EPSI NMR spectra showing the spectral dimension recorded at different spectral excitation bandwidths centered at the methyl region. A static (*i.e.*, non-flowing) and non-reacting mixture of BAEE and ethanol was used.

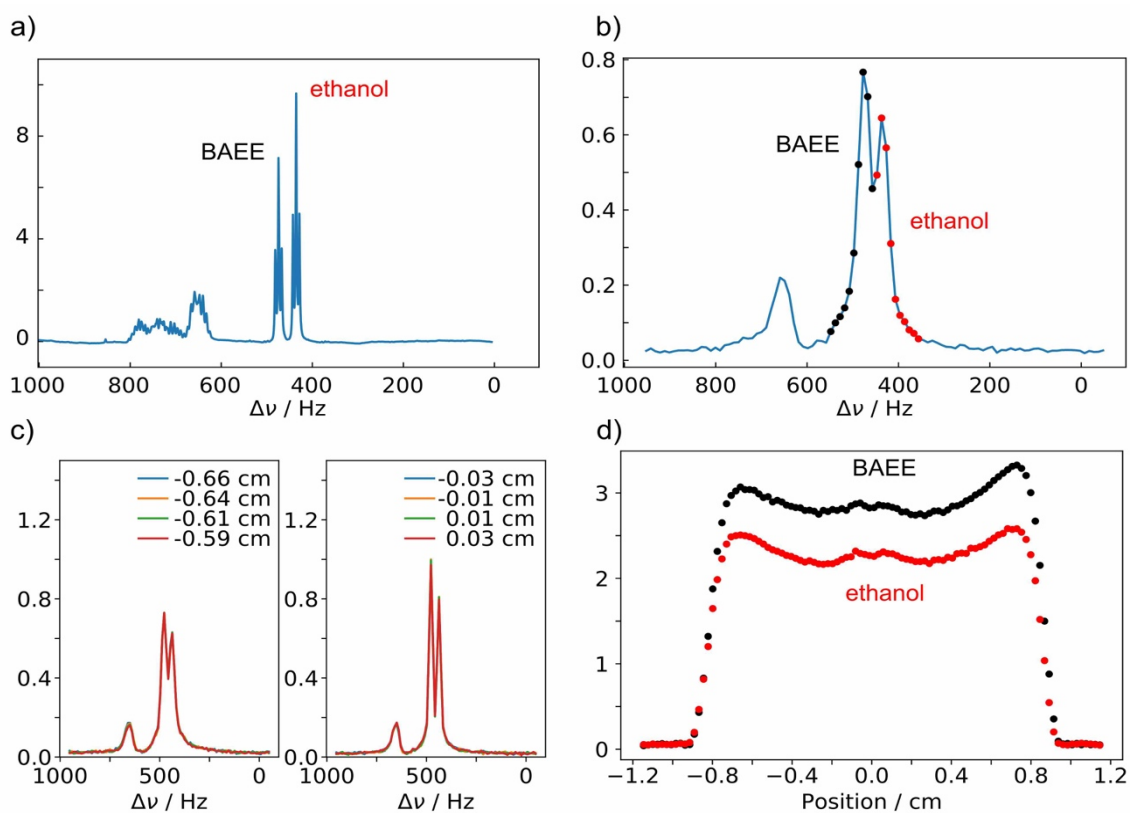

**Figure S.3**. **Integrating the $^1$H EPSI NMR reactant and product peaks for kinetic analysis.** a) 1D $^1$H NMR spectrum of a mixture of BAEE and ethanol. The integral ratio of ethanol over BAEE, $\alpha$, is larger than 1. b) Integration regions, indicated by the circles, for the reactant (black) and product (red). c) Spectral cross sections taken from two distinct regions along the field-of-view: from the edge of the coil (left) and in the middle of the coil (right). d) The integrals of BAEE ($I_R$) and ethanol ($I_P$) are plotted against position, which were further corrected to match the relative intensities measured in (a). Prior to the kinetic analysis the integrals IR and IP for each position were normalized according to $\hat{S}_R = S_R/(S_R + S_P)$ and $\hat{S}_P = S_P/(S_R + S_P)$. Only the signals contained with ± 0.7 cm of the center of the coil were used in the kinetic analysis, due inefficient excitation outside this region.

Supplementary Note 2. **The effect of the flow rate and the imaging spatial resolution on the signal-to-noise ratio.**

The effects of spatial resolution on the signal-to-noise ratio (SNR) of spectrotemporal 2D NMR correlations can be understood from the usual relationships between SNR and resolution in MRI.[1]

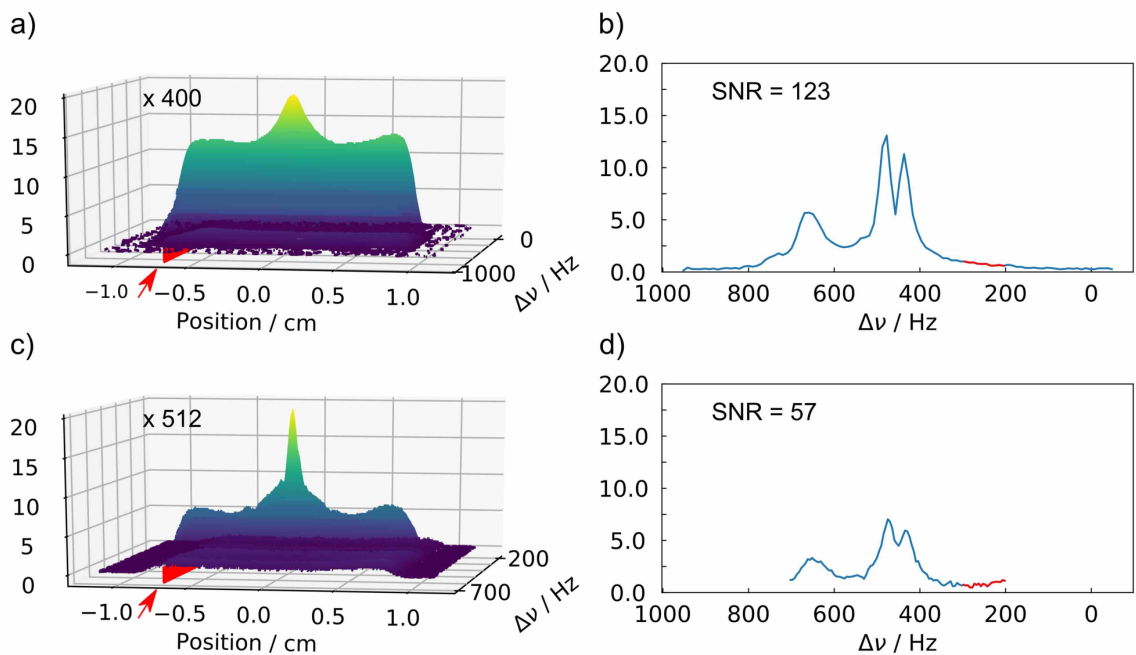

**Figure S.4. SNR measurements for $^1$H EPSI NMR spectra of non-flowing and non-reacting mixtures at different imaging spatial resolutions.** $^1$H EPSI NMR spectra comparing the SNR at different spatial resolutions (a-b, $\Delta z$ = 230 μm and c-d, $\Delta z$ = 115 μm) using a non-flowing and non-reacting sample mixture consisting of BAEE and ethanol. All other experimental gradient conditions are identical to those described in the main text. The red arrow in the 3D plots show the region ($z$ = 0.685 cm) where the SNR measurements are performed along the spatial dimension, whereas the red regions in the corresponding spectral dimension to the right indicate the noise regions that were considered. The SNR values are reported in the figure as well as the number of scans acquired in each case.

In this case, increasing the spatial resolution from *ca.* 230 μm to *ca.* 115 μm reduces the SNR by a factor of *ca.* 2, which is experimentally demonstrated for a non-reacting and non-flowing mixture of BAEE and ethanol acquired with two different spatial resolutions (**Figure S.4**) and is also confirmed by simulations (*vide infra*). Understanding the relative contributions of the flow rate on the SNR however, is more challenging as one must be able to accurately model the effects of flow when ideal plug-like conditions are no longer valid. To this end, experimental tests conducted at different flow rates demonstrate that the SNR decreases in a manner approximately commensurate with increasing flow rates (**Figure S.5** and **Figure S.6**). In this case, under the same experimental conditions, except at a flow rate that is 3.3× faster, the SNR decreases by a factor of *ca.* 2 (**Figure S.5a** *vs.* **Figure S.5c**). Simulations were conducted to investigate the relative contributions of spatial resolution and flow rate on the SNR (**Figure S.6**), which demonstrate that increasing flow rates do not result in SNR reductions. The SNR values for **Figures S.6a** with **S.6e**, which were simulated with identical gradient readout conditions ($\Delta z$ = 230 μm for both), but at two distinct non-zero flow rates ($v$ = 256 and 852 μm/s, respectively) are similar for each of the corresponding spectral cross sections (see **Table S1**). Furthermore, the SNR values measured for **Figure S.6c** and **Figure S.6g**, which were simulated to deliver higher spatial imaging resolution ($\Delta z = 115$ μm for both) are also similar for each of the corresponding spectral cross sections. Therefore, these simulations show that increasing the flow rate whilst holding all other acquisition parameters fixed does not result in significant changes to the SNR. However, if the flow rate and dead time ($t_0$) are both fixed and the imaging spatial resolution is doubled, then the simulations demonstrate that the SNR decreases by a factor of *ca*. 2, which agrees with the experimental results (**Figure S.6a** *vs.* **S.6c** and **S.6e** *vs.* **S.6g**). Modeling the effects of increasing reaction rate and dead time also results in an agreement with the experiment (**Figure S.6a** *vs.* **S.6b**, **S.6a** *vs.* **S.6b**, **S.6a** *vs.* **S.6b**, and **S.6a** *vs.* **S.6b**). This shows that faster flow rates must be used when targeting faster reaction rates in order to adequately "digitize" the spatiotemporally encoded dynamics.

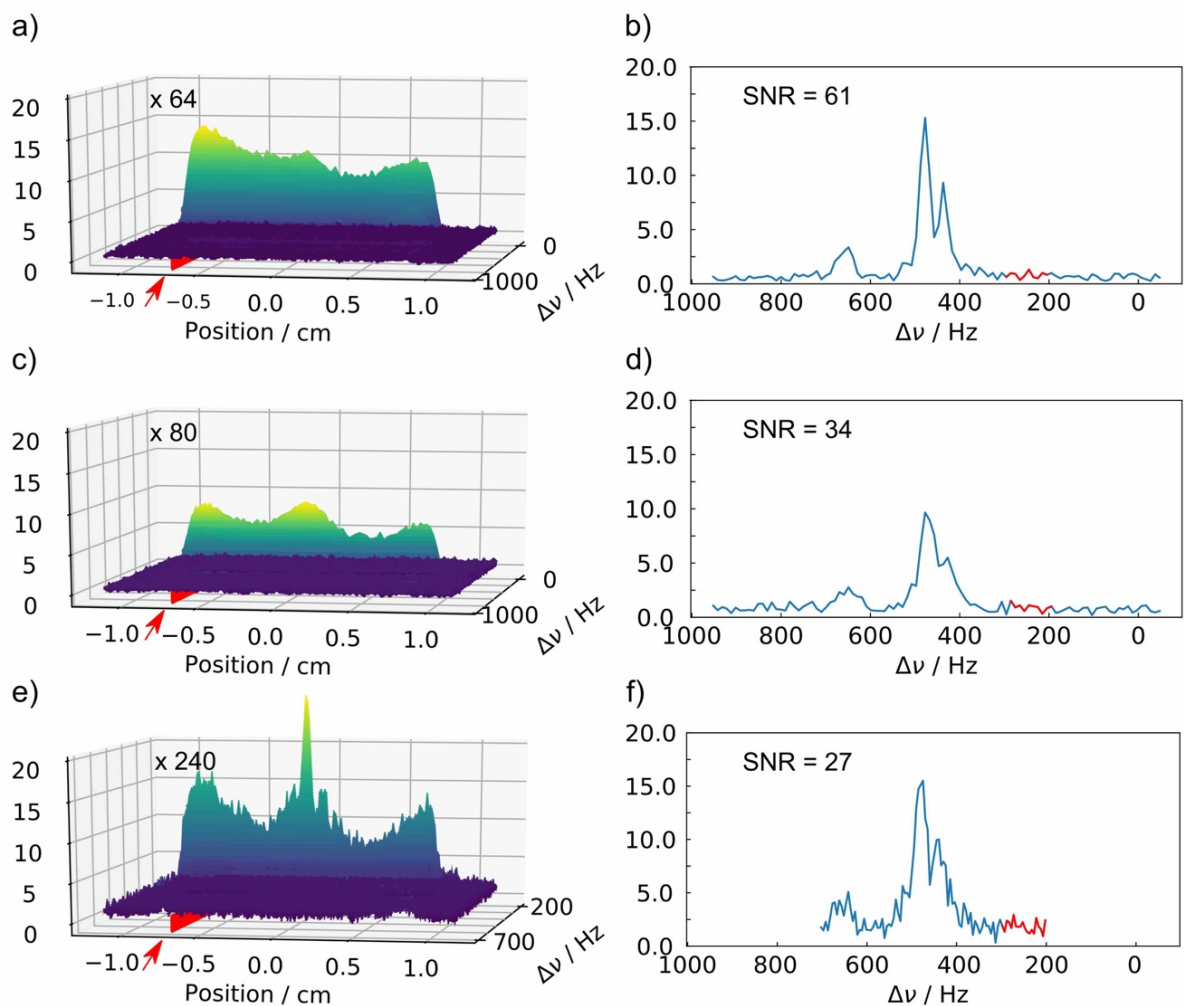

**Figure S.5. SNR measurements for $^1$H EPSI NMR spectra of non-reacting mixtures at different imaging spatial resolutions and flow rates.** SNR comparisons for $^1$H EPSI NMR spectra collected under various flow and gradient-readout conditions for the same non-reacting BAEE and ethanol mixture used in **Figure S.4**. The 3D view of the EPSI NMR spectra are shown in the left column and their corresponding spectral slices taken at the indicated spatial positions ($z = 0.685$ cm as indicated by the red arrows) are displayed to the right. The red regions in the $^1$H spectra in the right column indicate the noise regions (which were all the same) used for SNR measurements. The spatial resolution for the datasets presented in (a,b) and (c,d) is $\Delta z = 230$ μm whereas for (e,f) $\Delta z = 115$ μm. These were again achieved by appropriately adjusting the length of the readout gradient as described in the **Methods section**. The flow rate is $v = 0.3$ mL/min and $v = 1$ mL/min for (a-b) and (c-f), respectively. The SNR values are reported in the figure, as well as the number of scans used in each case (left column). For experiments conducted with the faster flow rate, the initial concentration of the enzyme was increased to $[E]_0 = 25$ μm.

**Table S1: SNR Values Measured from Numerically Simulated 1D Time-Resolved NMR Spectra (Figure S.6)**

| Position (cm) \ Figure S.6 | (a) | (b) | (c) | (d) | (e) | (f) | (g) | (h) |
|---|---|---|---|---|---|---|---|---|
| **0.7 (top)** | 30.1 | 52.0 | 12.7 | 3.0 | 25.8 | 7.9 | 12.8 | 4.8 |
| **0 (middle)** | 42.8 | 41.0 | 19.1 | 8.2 | 42.9 | 22.5 | 19.2 | 10.3 |
| **−0.7 (bottom)** | 53.6 | 31.1 | 32.8 | 19.7 | 53.1 | 34.9 | 32.4 | 21.9 |

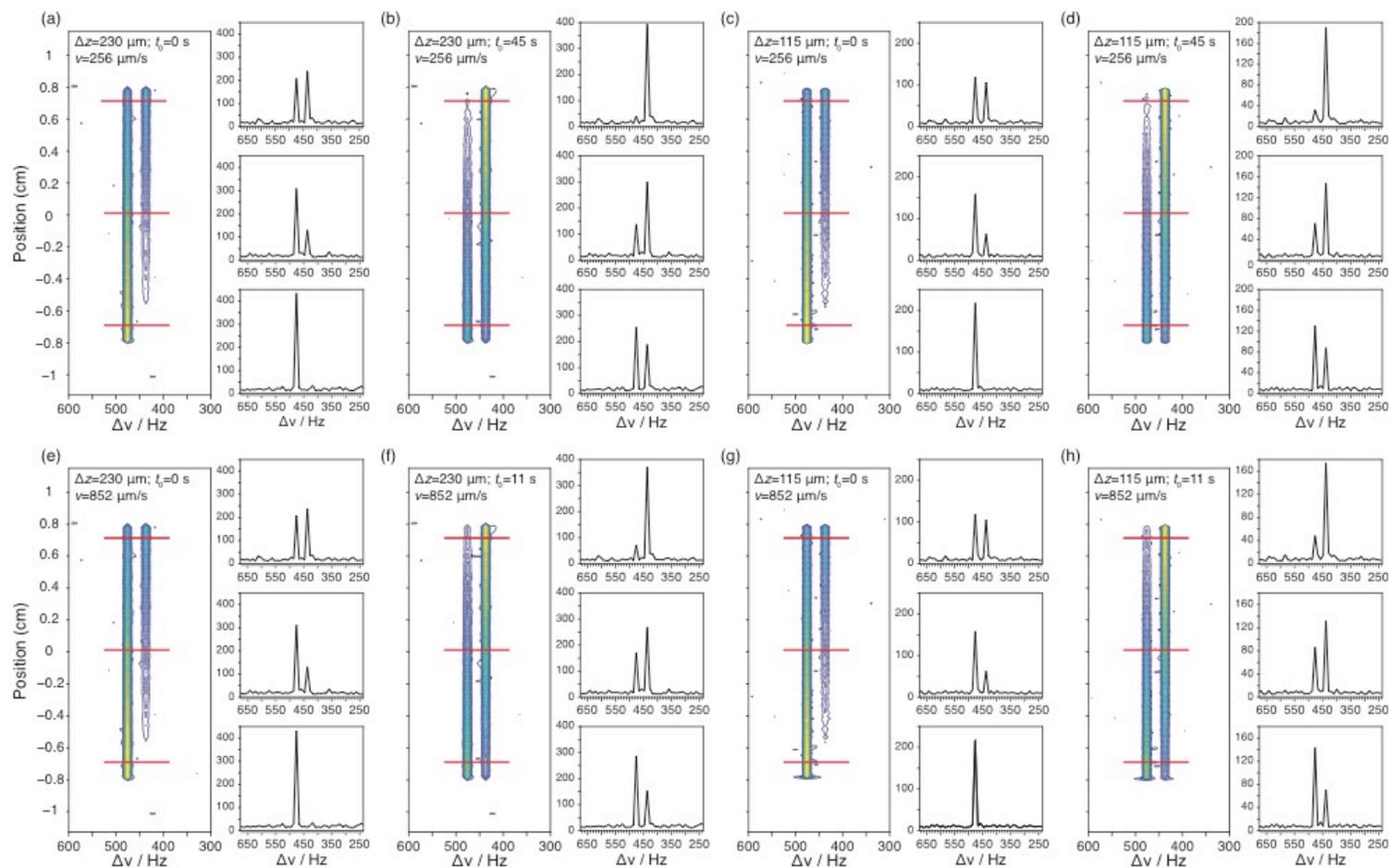

**Figure S.6. Numerical simulations on the effects of the imaging spatial resolution, dead time, and flow rate, on the SNR of 2D spectrotemporal correlations.** Numerical simulations of 2D $^1$H EPSI NMR spectra of the flowing off-equilibrium binary chemical reaction between BAEE and trypsin (monitoring the methyl regions as a function of time/position). These data show the combined effects of the spatial resolution, the linear flow rate ($v$), the dead time (*i.e.*, $t_0$), and the reaction conditions (*i.e.*, the initial concentration of reactants) on the overall SNR. All of these parameters, except for the concentrations, are indicated in each figure. The parameters used in these simulations mirror their experimental counterparts as closely as possible. The horizontal red bars indicate the positions from which the 1D time-resolved NMR spectra are extracted, which are displayed to the right of their corresponding 2D contour.

**Figure S.6 Cont.** The NMR spectra displayed in the top row were simulated with the same linear flow rate of $v = 256$ μm/s, whereas the NMR spectra displayed in the bottom row were simulated with $v = 852$ μm/s (*i.e.*, *ca.* 3.3× as fast). The NMR spectra displayed in the first two left-most columns (a, b, e, f) were simulated with the same spatial resolution ($\Delta z = 230$ μm) and the NMR spectra in the two right-most columns (c, d, g, h) were simulated under higher-resolution conditions ($\Delta z = 115$ um). It is noted that the gradient timing was doubled in order to increase the spatial resolution whilst other gradient parameters were held fixed. The NMR spectra occupying odd columns (*i.e.*, a, c, e, g) were simulated with $t_0 = 0$ s, whereas the datasets occupying the even columns (*i.e.*, b, d, f, h) were simulated with $t_0$ values matching their experimental counterparts (*i.e.*, $t_0 = 11$ s and 45 s). Lastly, the initial concentration of the enzyme trypsin was increased from 7.5 μM to 25 μM for the faster flow rate ($v = 852$ μm/s, smaller dead time) in order to accurately capture the zero-order rate kinetics in the sensitive region of the NMR coil (matching the conditions of the experiment). White normally-distributed Gaussian noise was added to each dataset. The corresponding SNR values are recorded in **Table S1**. All other simulation details are as discussed in the **Methods** section of the main text.

## Supplementary Note 3. Theoretical 1D NMR Spectra of Off-Equilibrium Irreversible Zero-Order Chemical Kinetics in the Absence of Flow and Field Gradients.

This paragraph provides a description for the NMR response of a reacting off-equilibrium chemical reaction characterized by a zero-order rate law (*i.e.*, $A \xrightarrow{k} B$), leading to the corresponding theoretical 1D NMR line shapes. The resulting model serves as a convenient starting point for elucidating the combined effects of plug-like flow and externally applied oscillating magnetic field gradients, which is discussed in the main text and elaborated upon in **Supplementary Note 4**. The reactant and product species are taken to be uniquely characterized by a single chemical environment. Each one is described by an isotropic chemical shift, whilst ignoring the effects of all other internal magnetic interactions. It is assumed here that initially only the reactant species **A** is present, and that its pre-polarized longitudinal equilibrium magnetization $M_z^A(t = 0)$ is proportional to its concentration $[A(t = 0)] = A_0$ for all positions $z$ within the NMR coil. Then, at $t = 0$ the **A** spin magnetization is instantaneously and homogeneously excited to the transverse plane and the chemical transformation is simultaneously initiated, which results in the progressive formation of the product species **B** – at the expense of the depletion of **A** – over the course of the FID acquisition. Ignoring the effects of relaxation for the moment, the change in the complex transverse magnetization for both **A** and **B** (denoted as $M_+^A(t)$ and $M_+^B(t)$, respectively) post excitation depends on their zero-order kinetics, in which the former and the latter decrease and increase linearly with time, respectively. The reaction progresses up to a point when all of **A** is expended (**Figure S.7a**). At this point in time, which we denote as $t_{\max}$, the complex transverse magnetization of **A** (as well as its concentration) is 0, and remain so for all $t \geq t_{max}$. Concomitantly, the absolute magnitude of the transverse magnetization for **B** becomes equal to $A_0$ for all $t \geq t_{max}$. The **A** spin NMR signal is relatively straightforward to model. It is the product between its time-dependent complex transverse magnetization and its chemical shift evolution (accounting for the effects of transverse relaxation $T_2$), as given by:

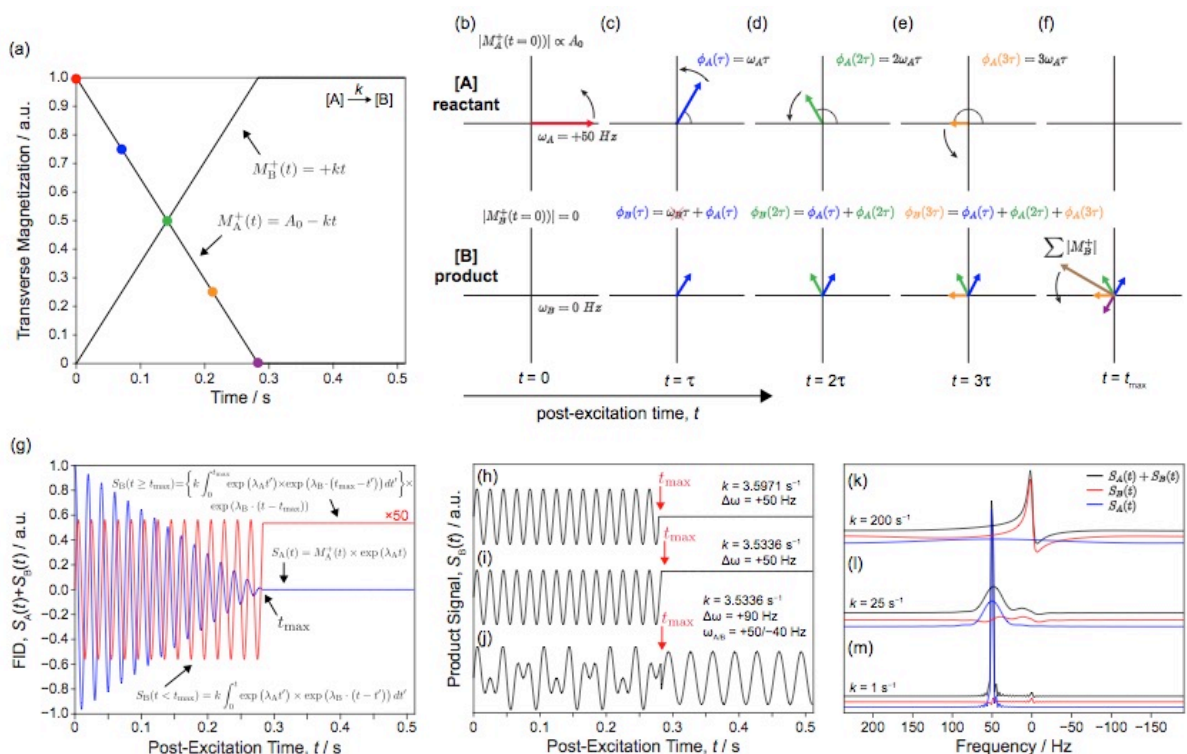

**Figure S.7. Time-dependent transverse magnetization dynamics of A and B evolving from zero-order chemical kinetics occurring over the course of the FID acquisition.** (a) The coloured points indicate distinct time points throughout the reaction process and correspond with the vector model presented in (b)-(f), which schematically illustrates the dynamics of **A** and **B** in a rotating reference frame that is resonant with the chemical shift of **B**. In this case, **A** oscillates with a +50 Hz frequency, whereas the magnetization vector elements of **B**, which are continuously formed throughout the reaction process, remain stationary. (g) FID evolution of **A** (blue trace) and **B** (red trace, × 50) for $\omega_A = +50$ Hz, $\omega_B = 0$ Hz, and $k = 3.5336$ s$^{-1}$. (h)-(j) Illustrations of the **B** signal evolution for different exchange rates and chemical shift separations (values indicated in the figure). (k)-(m) 1D NMR line shapes simulated at different exchange rates (values indicated in the figure) for $\omega_A = +50$ Hz and $\omega_B = 0$ Hz.

$$S_A(t) \propto \begin{cases} M_+^A(t) \times \exp(\lambda_A t) = (A_0 - kt) \times \exp(\lambda_A t), & t < t_{max} \\ 0, & t \geq t_{max} \end{cases} \quad (S1.1)$$

where $A_0$ is the initial concentration of **A** in units of M (mol/L), $k$ is the zero-order reaction rate in units of M s$^{-1}$ and $\lambda_A = i\omega_A - R_{2A}$, where $\omega_A$ and $R_{2A}$ are the **A**-spin chemical shift in rad/s and transverse relaxation rate in s$^{-1}$, respectively. This is Equation (2a) in the main text.

Describing the time evolution of **B** needs to account that the entirety of **B** progressively forms from **A** over the course of the FID acquisition. These spins will therefore spend some finite amount of time $0 \leq t < t'$ evolving as **A** under the effects of $\omega_A$, $R_{2A}$, and the **A**-spin chemical kinetics, accumulating a complex exponent proportional to $\lambda_A t$. Then, at some time point $t = t'$, **B** is instantaneously formed with probability $P_{A \rightarrow B}(t') = k$. By then the spin packet has accrued an exponential factor equal to $\lambda_A t'$; subsequently it evolves under the effects of the **B**-spin characteristics ($\omega_B, R_{2B}$) for $t' \leq t < t_{max}$. This unitary transformation and subsequent transfer of phase information from **A** → **B** continues until all of **A** has reacted (*i.e.*, when $t = t_{max}$), at which point the **B** transverse complex magnetization – which is itself comprised of discrete $t'$-dependent **A**-spin phase components – evolves in unison at the chemical shift of **B** for all $t \geq t_{max}$. Mathematically, the **B** signal can thus be represented in terms of these two distinct time-dependent contributions as:

$$S_B(t < t_{max}) \propto \int_0^t P_{A \rightarrow B}(t') \times \exp(\lambda_A t') \times \exp\left(\lambda_B \cdot (t - t')\right) dt', \quad (S1.2)$$

and,

$$S_B(t \geq t_{max}) \propto \left\{ \int_0^{t_{max}} P_{A \rightarrow B}(t') \times \exp(\lambda_A t') \times \exp\left(\lambda_B \cdot (t_{max} - t')\right) dt' \right\} \times \exp\left(\lambda_B \cdot (t - t_{max})\right), \quad (S1.3)$$

where $P_{A \rightarrow B}(t') = k$ is the probability for the reaction to occur at $t'$ and $\lambda_B = i\omega_B - R_{2B}$, which has the same units as the $\lambda_A$ defined above. These integrals readily evaluate to the following expressions to give the signal of **B** for all post-excitation times $t$:

$$S_B(t < t_{max}) \propto k \exp(\lambda_B t) \int_0^t \exp\left((\lambda_A - \lambda_B) t'\right) dt'$$

$$S_B(t < t_{max}) \propto \left(\frac{k}{\lambda_A - \lambda_B}\right) \times \exp(\lambda_B t) \times \left\{\exp\left((\lambda_A - \lambda_B) t'\right)\right\}\Big|_{t'=0}^{t'=t}$$

$$S_B(t < t_{max}) \propto \left(\frac{k}{\lambda_A - \lambda_B}\right) \times (\exp(\lambda_A t) - \exp(\lambda_B t)) \quad (S1.4)$$

and,

$$S_B(t \geq t_{max}) \propto k \exp(\lambda_B t_{max}) \times \left\{ \int_0^{t_{max}} \exp((\lambda_A - \lambda_B) t') dt' \right\}$$
$$\times \exp(\lambda_B \cdot (t - t_{max}))$$

$$S_B(t \geq t_{max}) \propto \left( \frac{k}{\lambda_A - \lambda_B} \right) \times \exp(\lambda_B t_{max}) \times \left\{ \exp((\lambda_A - \lambda_B) t') \right\} \Big|_{t'=0}^{t'=t_{max}}$$
$$\times \exp(\lambda_B \cdot (t - t_{max}))$$

$$S_B(t \geq t_{max}) \propto \left( \frac{k}{\lambda_A - \lambda_B} \right) \times (\exp(\lambda_A t_{max}) - \exp(\lambda_B t_{max}))$$
$$\times \exp(\lambda_B \cdot (t - t_{max})) \tag{S1.5}$$

Combining Equations (S1.4) and (S1.5) gives the piece-wise expression for the evolution of the **B**-spin NMR signal:

$$S_B(t) \propto \begin{cases} \left( \frac{k}{\lambda_A - \lambda_B} \right) \times (\exp(\lambda_A t) - \exp(\lambda_B t)), & t < t_{max} \\ \left( k \frac{(\exp(\lambda_A t_{max}) - \exp(\lambda_B t_{max}))}{\lambda_A - \lambda_B} \right) \times \exp(\lambda_B \cdot (t - t_{max})), & t \geq t_{max}. \end{cases} \tag{S1.6}$$

This is Equation (2b) in the main text. Contributions to the **B**-spin NMR signal arising from excited equilibrium magnetization present at $t = 0$ are straight-forward to account for, and are described by a complex exponential akin to that in Equation (S1.1), evolving at $\lambda_B$ and possessing a pre-exponential magnetization value that is time independent (*i.e.*, independent of the chemical kinetics).

The spin dynamics described by Equations (S1.1) and (S1.6) can be represented with a simplistic vector model in which the evolution of the bulk transverse magnetization for both the reactant and the product are independently examined in the same rotating reference frame (**Figure S.7b-S.7f**). This description can be further simplified by assuming that the transmitter frequency is resonant with $\omega_B$, and therefore the chemical shift evolution of the product NMR signal is conveniently equal to zero. The plots in **Figure S.7b-S.7f** display snapshots of the evolution of the magnetization for **A** and **B** (top and bottom, respectively) at regularly spaced $\tau$ intervals throughout the reaction/acquisition, which are designated by the distinct colours that correspond to the coloured sampling points along the reaction coordinate as indicated in **Figure S.7a**. Initially, at $t = 0$ the **A**-spin magnetization is excited along the +*x* axis of the rotating frame (**Figure S.7b**, red vector), which in this example begins to evolve with a +50 Hz chemical shift – there is no measurable **B** magnetization at $t = 0$. At an instant in time $t = \tau$ post excitation, the **A** spin transverse magnetization acquires a phase proportional to $\omega_A \tau$, whilst simultaneously giving rise to a discrete transverse **B** magnetization vector that possesses the same **A**-spin phase and whose magnitude is proportional to $k\tau$ (**Figure S.7c**). Immediately after this instant of time, the **A**-spin magnetization continues to precess at +50 Hz and decays according to the zero-order kinetics (this is schematically indicated by a shortening of the **A** vector magnitude as described by Equation (S1.1)), whilst the component of **B** created precisely at $t = \tau$ remains stationary with a discrete phase given solely by $\phi_A(\tau)$ (*i.e.*, since $\omega_B = 0$ Hz and the newly formed **B** magnetization element does not accumulate any additional phase when the transmitter is on resonance with the product peak), and is affected only by $T_2$ relaxation. At a later time, $t = 2\tau$, another discrete packet of **B** magnetization is formed that again has a magnitude proportional to $k\tau$, but at this moment possesses a phase equal to $2\omega_A \tau$ that continues to remain stationary for all $t$ (**Figure S.7d**, bottom row). This behaviour continues for all time points $t < t_{\max}$ until the reaction is completed (**Figure S.7e**). It is important to stress that **B** continuously and progressively forms from **A** for all $t < t_{\max}$, with the overall transverse magnetization of **B** comprised of discrete and individual vector elements that all have the same magnitude, but have distinct and unique phase contributions that are given by the instantaneous **A**-spin phase that corresponds to the state of **A** at the time a given **B** magnetization vector element was formed. The total **B** signal is therefore given by the sum of all the individual **B** vector elements that are formed for all infinitesimal time elements $dt$ when $t < t_{\max}$, which is mathematically represented by the integral given by Equation (S1.2). At $t = t_{\max}$, when all of **A** has reacted, the entire ensemble of discrete **B** magnetization vector elements then evolves in unison (**Figure S.7f**) under the effects of the **B**-spin Liouvillian and precesses collectively at $\omega_B$, which for this example is zero (*i.e.*, on resonance with the **B** peak), thus each of the individual **B** vector elements remain stationary in the rotating frame (Equation (S1.3)). Plots of these time-domain dynamics (in the absence of $T_2$ relaxation) are displayed in **Figures S.7g-S.7j**, which show a simultaneous decay (*i.e.*, resulting from the irreversible chemical kinetics) and +50 Hz chemical shift oscillation of **A** for $t < t_{\max}$ (**Figure S.7g**, blue trace), as well as the corresponding $\Delta\omega_{AB} = \omega_A - \omega_B = +50 - 0 = +50$ Hz oscillation of **B** when $t < t_{\max}$ (**Figure S.7g**, red trace). At times $t \geq t_{\max}$ post excitation, the irreversible reaction has gone to completion as is indicated by the absence of the **A** signal (**Figure S.7g**, blue trace) and the stationary non-oscillatory **B** signal. It is important to note that the integral in Equation (S1.3) sets the initial phase of the subsequent $\omega_B$ oscillation when $t \geq t_{max}$. This point is illustrated further in **Figure S.7h** and **Figure S.7i,** which were simulated for the same $\Delta\omega_{AB}$ frequency difference of +50 Hz, but at two slightly different $k$ rates (3.5971 s$^{-1}$ and 3.5336 s$^{-1}$), resulting in $\omega_B$ trajectories that differ by 90°. **Figure S.7j** shows the behaviour of the signal of **B** for a non-zero $\omega_B$ chemical shift ($\omega_A$ = +50 Hz and $\omega_B$ = −40 Hz), which results in a multi-frequency oscillation for $t < t_{max}$ that eventually evolves into a single frequency oscillation at the $\omega_B$ frequency when all of **A** has reacted. **Figures S.7k-S.7m** showcase the typical line shapes expected from these off-equilibrium zero-order NMR kinetics, which were simulated in the fast (**Figure S.7k**), intermediate (**Figure S.7l**), and slow (**Figure S.7m**) exchange regimes. This model is now the starting point for elucidating the effects of plug-like flow before and during FID acquisition, which has the effect of spatiotemporally encoding the reaction kinetics along the spatial coordinate of the NMR coil, whose effects are discussed in the main text.

Supplementary Note 4. **Flowing spins in oscillating magnetic field gradients**.

The following analysis investigates the spin dynamics arising from non-reacting NMR-active nuclei that are simultaneously affected by uniform plug-like flow and different magnetic field gradient waveforms. The combined effects of motion and field gradients on the resulting NMR responses are well understood;[2–4] several examples that are relevant to the EPSI-based kinetic NMR method discussed in the main text, are recapitulated here for the sake of clarity. In all cases, the spin dynamics evolving under the gradient waveforms shown in **Figure S.8** and elsewhere in this Paragraph, are derived assuming plug-like flow conditions characterized by a uniform flow velocity *v*.

*(i) A single bipolar gradient phase encodes flow and echoes the effects of stationary spins*

**Figure S.8a** shows a single bipolar readout gradient that is applied in the same direction as the uniform flow, which is in the *z* direction. The phase evolution accrued for a single spin at an initial position $z_0$ is determined for each gradient lobe as

$$\phi_1(0 \leq t \leq T_a) = \int_0^{T_a} \gamma G_z(t')z(t')dt' = +G_z\gamma \int_0^{T_a} (z_0 + vt')dt' = +G_z\gamma \left\{z_0[T_a] + \frac{1}{2}v[T_a^2]\right\} \quad (S2.1)$$

$$\phi_2(T_a \leq t \leq 2T_a) = \int_{T_a}^{2T_a} \gamma G_z(t')z(t')dt' \quad (S2.2)$$

$$= -G_z\gamma \int_{T_a}^{2T_a} (z_0 + vt')dt' = -G_z\gamma \left\{z_0[2T_a - T_a] + \frac{1}{2}v[4T_a^2 - T_a^2]\right\}$$

$$= -G_z\gamma \left\{z_0[T_a] + \frac{3}{2}v[T_a^2]\right\}$$

$$\phi_1(0 \leq t \leq T_a) + \phi_2(tT_a \leq t \leq 2T_a) = G_z\gamma \left\{z_0[0] + v\left[\frac{1}{2}T_a^2 - \frac{3}{2}T_a^2\right]\right\} = -\gamma G_z v T_a^2 \quad (S2.3)$$

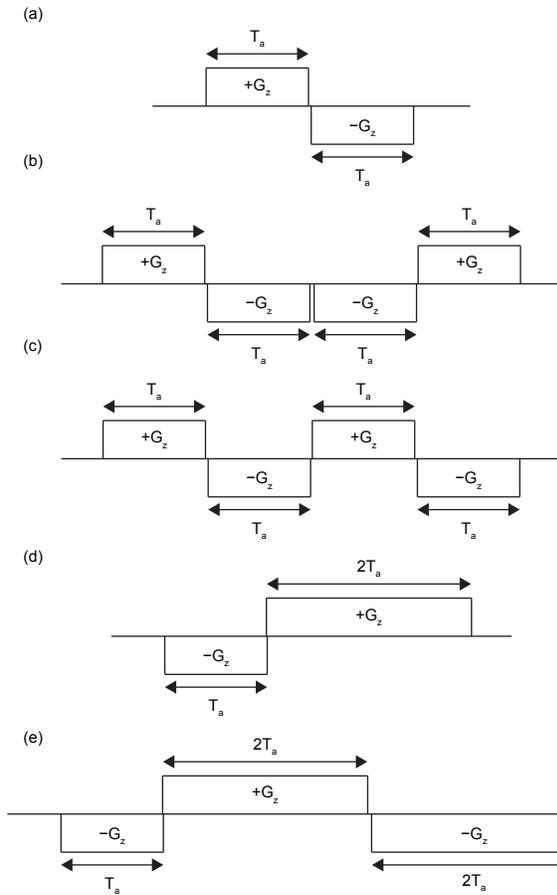

**Figure S.8. Evaluating the effects of various gradient waveforms and uniform flow on the NMR signal phase.** Schematic representations of the gradient sequences over which the transverse phase is calculated for a single $z_0$ position under the effects of uniform flow. As show in the text variants (a), (c), and (d) are not flow compensated – the transverse phase is not simultaneously refocused from contributions originating from static spins and moving spins. By contrast (b) and (e) are flow compensated.

Therefore, Equation (S2.3) demonstrates that at the end of the second gradient pulse, the contribution of stationary spins (*i.e.*, $v$ independent spins) to the transverse phase is refocused whilst the contributions emanating from moving spins remain. This type of bipolar gradient readout is therefore selective for spins that are moving after a duration of $2T_a$.

*(ii) A symmetric pair of bipolar gradients is flow-compensated*

**Figure S.8b** shows a pair of symmetrically-placed bipolar gradients, that are again applied along the z direction collinear with the uniform flow. $\phi_1$ and $\phi_2$ are the same in this case as derived above, so $\phi_3$ and $\phi_4$ are then:

$$\phi_3(2T_a \leq t \leq 3T_a) = \int_{2T_a}^{3T_a} \gamma G_z(t')z(t')dt'$$

$$= -G_z\gamma \int_{2T_a}^{3T_a} (z_0 + vt')dt'$$

$$= -G_z\gamma \left\{z_0[3T_a - 2T_a] + \frac{1}{2}v[9T_a^2 - 4T_a]\right\}$$

$$= -G_z\gamma \left\{z_0[T_a] + \frac{5}{2}v[T_a^2]\right\} \tag{S2.4}$$

$$\phi_4(3T_a \leq t \leq 4T_a) = \int_{3T_a}^{4T_a} \gamma G_z(t')z(t')dt'$$

$$= +G_z\gamma \int_{3T_a}^{4T_a} (z_0 + vt')dt'$$

$$= +G_z\gamma \left\{z_0[4T_a - 3T_a] + \frac{1}{2}v[16T_a^2 - 9T_a]\right\}$$

$$= +G_z\gamma \left\{z_0[T_a] + \frac{7}{2}v[T_a^2]\right\} \tag{S2.5}$$

All together now,

$$\phi_1(0 \leq t \leq T_a) + \phi_2(T_a \leq t \leq 2T_a) + \phi_3(2T_a \leq t \leq 3T_a)$$
$$+ \phi_4(t3T_a \leq t \leq 4T_a)$$
$$= -\gamma G_z v T_a^2 + 0 + G_z\gamma\left\{-\frac{5}{2}v[T_a^2] + \frac{7}{2}v[T_a^2]\right\}$$
$$= -\gamma G_z v T_a^2 + \gamma G_z v T_a^2 = 0 \tag{S2.6}$$

Therefore, the total transverse phase after 2 bipolar readouts (**Figure S.8b**) is equal to zero for both stationary spins and spins moving with uniform velocity.

*(iii) A pair of bipolar gradients is not flow-compensated*

**Figure S.8c** shows a pair of bipolar readout gradients, of the kind that would concatenate in an EPSI sequence. The phase accumulated at $\phi_3$ and $\phi_4$ in this case is:

$$\phi_3(2T_a \leq t \leq 3T_a) = \int_{2T_a}^{3T_a} \gamma G_z(t')z(t')dt'$$

$$= +G_z\gamma \int_{2T_a}^{3T_a} (z_0 + vt')dt'$$

$$= +G_z\gamma \left\{z_0[3T_a - 2T_a] + \frac{1}{2}v[9T_a^2 - 4T_a^2]\right\}$$

$$= +G_z\gamma \left\{z_0[T_a] + \frac{5}{2}v[T_a^2]\right\} \tag{S2.7}$$

$$\phi_4(3T_a \leq t \leq 4T_a) = \int_{3T_a}^{4T_a} \gamma G_z(t')z(t')dt'$$

$$= -G_z\gamma \int_{3T_a}^{4T_a} (z_0 + vt')dt'$$

$$= -G_z\gamma \left\{z_0[4T_a - 3T_a] + \frac{1}{2}v[16T_a^2 - 9T_a]\right\}$$

$$= -G_z\gamma \left\{z_0[T_a] + \frac{7}{2}v[T_a^2]\right\} \tag{S2.8}$$

Adding $\phi_1$ to $\phi_4$ gives:

$$\phi_1(0 \leq t \leq T_a) + \phi_2(T_a \leq t \leq 2T_a) + \phi_3(2T_a \leq t \leq 3T_a)$$
$$+ \phi_4(t3T_a \leq t \leq 4T_a)$$
$$= -\gamma G_z v T_a^2 + 0 + G_z\gamma \left\{+\frac{5}{2}v[T_a^2] - \frac{7}{2}v[T_a^2]\right\}$$
$$= -\gamma G_z v T_a^2 - \gamma G_z v T_a^2 = -2\gamma G_z v T_a^2 \tag{S2.9}$$

Equation (S2.9) shows that a pair of bipolar gradients is not flow-compensated: it refocuses the phase evolution originating from static spins, but moving spins accrue twice the phase as in a single bipolar gradient.

*(iv) A gradient echo is not flow compensated*

**Figure S.8d** shows a conventional gradient echo sequence with the pre-phase and readout gradient applied along the flow direction, $z$. The transverse phase accumulated under the action of the pre-phaser, $\phi_1$, is:

$$\phi_1(0 \leq t \leq T_a) = \int_0^{T_a} \gamma G_z(t')z(t')dt' = -\gamma G_z \int_0^{T_a} (z_0 + vt')dt' = -\gamma G_z \left\{ z_0[T_a] + \frac{1}{2}v[T_a^2] \right\} \quad (S2.7)$$

Similarly, the phase accumulated up to the middle of $\phi_2$ (*i.e.*, at the point the stationary spins are refocused) is:

$$\phi_2(T_a \leq t \leq 2T_a) = \int_{T_a}^{2T_a} \gamma G_z(t')z(t')dt'$$
$$= +\gamma G_z \int_{T_a}^{2T_a} (z_0 + vt')dt' = +\gamma G_z \left\{ z_0[2T_a - T_a] + \frac{1}{2}v[4T_a^2 - T_a^2] \right\}$$
$$= +\gamma G_z \left\{ z_0[T_a] + \frac{1}{2}v[3T_a^2] \right\} \quad (S2.8)$$

It's clear that although stationary spins produce an echo, the phase contributions from flowing spins are not refocused in the center of the gradient echo. Continuing this analysis for the rest of the reversed gradient, the phase at the end of this gradient is:

$$\phi_2(T_a \leq t \leq 3T_a) = \int_{T_a}^{3T_a} \gamma G_z(t')z(t')dt' \quad (S2.9)$$
$$= +\gamma G_z \int_{T_a}^{3T_a} (z_0 + vt')dt' = +\gamma G_z \left\{ z_0[3T_a - T_a] + \frac{1}{2}v[9T_a^2 - T_a^2] \right\}$$
$$= +\gamma G_z \{ z_0[2T_a] + v[4T_a^2] \}$$

The total phase at the end of the sequence is therefore

$$\phi_1(0 \leq t \leq T_a) + \phi_2(T_a \leq t \leq 3T_a) = -\gamma G_z \left\{ z_0[T_a] + \frac{1}{2}v[T_a^2] \right\} + \gamma G_z\{z_0[2T_a] + v[4T_a^2]\} \quad (S2.10)$$
$$= +\gamma G_z \left\{ z_0[T_a] + \frac{7}{2}v[T_a^2] \right\}$$

*(v) A pre-phased bipolar gradient is flow compensated*

**Figure S.8e** shows a bipolar gradient readout preceded by a pre-phasing period lasting half the length of a single readout gradient (*i.e.*, $T_a$). In this case, only $\phi_3$ needs to be explicitly calculated, since the phase evolution for the first read gradient and pre-phaser is identical to that of a conventional gradient echo (Equation (S2.13)). Therefore,

$$\phi_3(3T_a \leq t \leq 4T_a) = \int_{3T_a}^{4T_a} \gamma G_z(t')z(t')dt' \quad (S2.14)$$
$$= -\gamma G_z \int_{3T_a}^{4T_a} (z_0 + vt')dt' = -\gamma G_z \left\{ z_0[4T_a - 3T_a] + \frac{1}{2}v[16T_a^2 - 9T_a^2] \right\}$$
$$= -\gamma G_z \left\{ z_0[T_a] + \frac{7}{2}v[T_a^2] \right\}$$

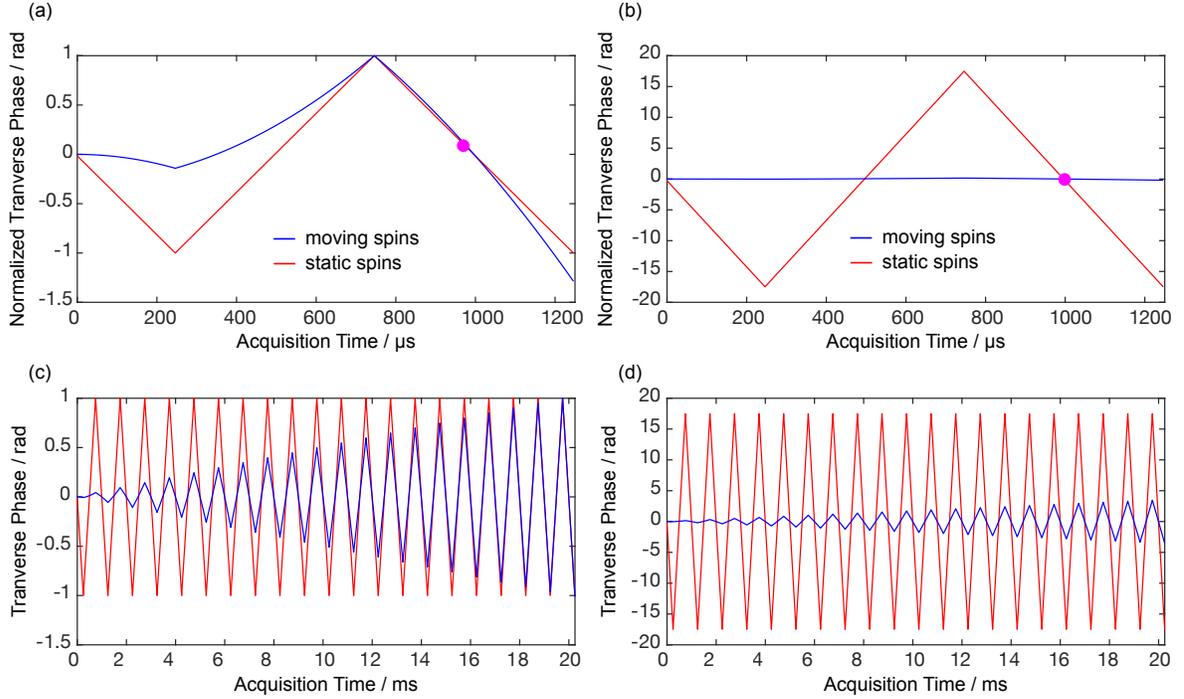

**Figure S.9.** Plots of the transverse phase resulting from the action of a pre-phased bipolar readout gradient (shown in **Figure S.8e**, which is the basis of an EPSI readout with $N_{EPSI} = 1$). Plots in the top row show the phase from the static (red) and moving spins (blue) over the course of the three gradients (prephase - (+)-ve gradient - (-)-ve) with $T_a = 500$ μs and $G_z = 26.14$ gauss/cm. Plots in the bottom row show the phase evolution over $N_{EPSI} = 20$ readouts. Data in the left column are normalized, whereas data in the right column are unnormalized. The phase was calculated for $z_0 = 0.1$ cm and $v = 1$ cm/s.

$\phi_3$ is the transverse phase calculated until the gradient echo is formed in the center of the third gradient pulse. Adding Equation S2.13 to S2.14 gives:

$$\phi_1(0 \leq t \leq T_a) + \phi_2(T_a \leq t \leq 3T_a) + \phi_3(3T_a \leq t \leq 4T_a) \quad \text{(S2.11)}$$
$$= +\gamma G_z \left\{ z_0[T_a] + \frac{7}{2} v[T_a^2] \right\} - \gamma G_z \left\{ z_0[T_a] + \frac{7}{2} v[T_a^2] \right\} = 0$$

Equation (S2.15) shows that a bipolar gradient applied with a pre-phaser does indeed refocus the phase evolution of both the static and moving spins. The plots in **Figure S.9** show the evolution of both the phase for the static spins (shown in red) and spins with uniform velocity (shown in blue) over the course of one (**Figure S.9a** and **Figure S.9b**) and 20 (**Figure S.9c** and **Figure S9.d**) bipolar readout gradients with an initial phasing period lasting half the length of a single readout gradient. The plots in the left column are normalized with respect to the phase accumulated by the static spins in order to better compare their values to those of the moving spins, since the contributions to the phase emanating from the latter are small for small values of *t*. The magenta circles indicate the time values when the phase for static and moving spins are both equal to zero.

## Supplementary Note 5. **Gradient-driven retrieval of spatiotemporally-encoded zero-order kinetics: Derivation and analysis**.

In this section, the mathematical details concerning the derivation of Equations (5) of the main text are provided, in addition to a brief analysis of the expressions that describe the spin dynamics of time-resolved spatiotemporally-encoded kinetic NMR spectra. Composing the expression that describes the NMR evolution of product over the course of an EPSI readout begins by multiplying each unique chemical shift term in Equations (S1.2) and (S1.3) by its corresponding gradient-induced frequency shift, which gives for Equation (S1.2),

$$S_B(t < t_{max}) \propto \int_Z M_+^B(z, t) \times \exp(\lambda_B t) \times \exp\left(i\gamma \int_0^t G(t') \cdot z(t')dt'\right) dz$$
$$+ \int_Z \left\{ \int_0^t P_{A \to B}(t') \times \exp(\lambda_A t') \times \exp\left(i\gamma \int_0^{t'} G(t'') \cdot z(t'')dt''\right) \right.$$
$$\left. \times \exp(\lambda_B \cdot (t - t')) \times \exp\left(i\gamma \int_{t'}^t G(t'') \cdot z(t'')dt''\right) dt' \right\} dz \qquad (S3.1)$$

and for Equation (S1.3),

$$S_B(t \geq t_{max}) \propto \int_Z M_+^B(z, t) \times \exp(\lambda_B t) \times \exp\left(i\gamma \int_0^t G(t') \cdot z(t')dt'\right) dz$$
$$+ \int_Z \left[ \left\{ \int_0^{t_{max}} dt' P_{A \to B}(t') \times \exp(\lambda_A t') \times \exp\left(i\gamma \int_0^{t'} G(t'') \cdot z(t'')dt''\right) \right. \right.$$
$$\left. \times \exp(\lambda_B \cdot (t_{max} - t')) \times \exp\left(i\gamma \int_{t'}^{t_{max}} G(t'') \cdot z(t'')dt''\right) \right\}$$
$$\left. \times \exp(\lambda_B \cdot (t - t_{max})) \times \exp\left(i\gamma \int_{t_{max}}^t G(t'') \cdot z(t'')dt''\right) \right] dz, \qquad (S3.2)$$

Simplifying the last term in both equations representing the gradient-induced shifts by exploiting the properties of integrals that have common integration limits and then integrating over $t$ gives the piece-wise expression for the **B** signal contained in Equations (5b) and (5c) of the main text. These equations (5) can then be used to model the experimental kinetic EPSI NMR spectra, of which a representative example is shown in **Figure S.10** for three limiting cases. In **Figure S.10a**, the linear flow rate is set to 2 cm/s with a reaction rate $k = 0$ s$^{-1}$, which gives the EPSI NMR spectrum of only the reactant under conditions of continuous flow. The effects of the flow-induced broadening are clearly observed over spatial regions that eventually become vacated of NMR-emitting spins. **Figure S.10b** shows the EPSI NMR spectrum of the off-equilibrium reaction ($k = 1$ s$^{-1}$) acquired under stopped-flow conditions, leading to the absence of flow-induced broadening. The flow rate before acquisition was 2 cm/s, which results in the spatiotemporal encoding of the reaction kinetics. Lastly, in **Figure S.10c**, both the reaction rate and the flow rate are non-zero over the course of the EPSI readout, which correspond to experimental-like conditions ($v = 2$ cm/s and $k = 1$ s$^{-1}$). Notice how the same degree of flow-induced broadening affects the spectral dimension of the EPSI readout as in **Figure S.10a**. Furthermore, the contributions to the **B** signal that arise from reaction of **A** post-excitation are largely absent from these data due to the small value of $k$, leading to signal intensities that are orders of magnitude smaller than the signal intensity arising from equilibrium magnetization formed pre-excitation.

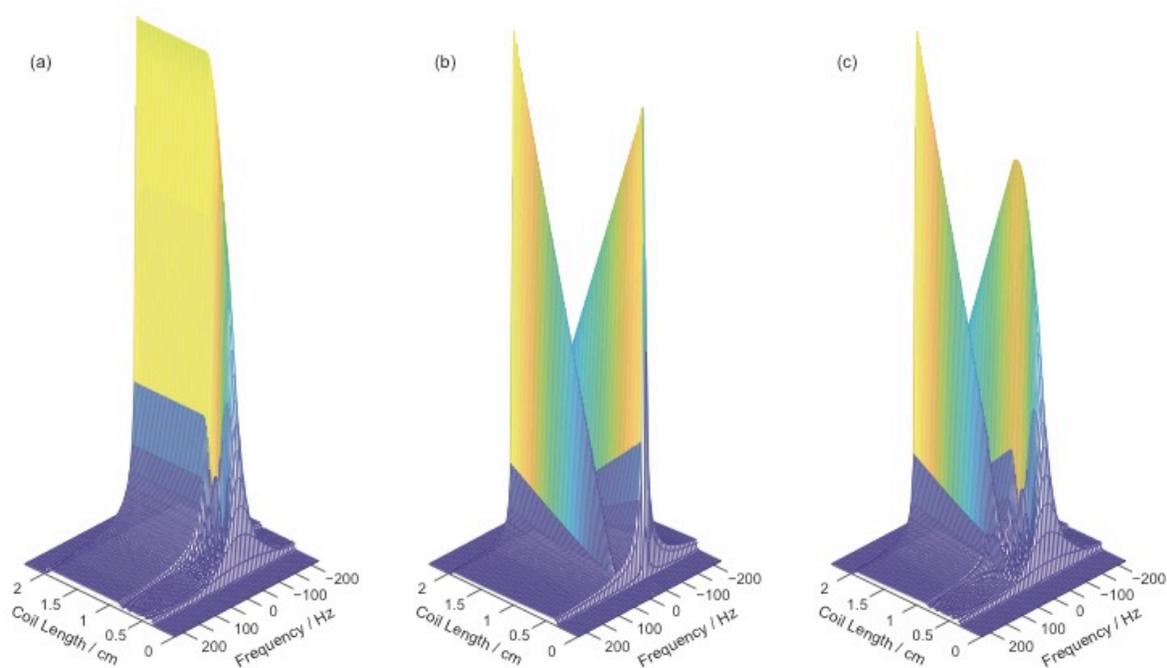

**Figure S.10**. **Analytical simulations of EPSI NMR spectra using Equations (7) and (8) of the main text.** Simulated EPSI NMR spectra assuming (a) $k = 0$ s$^{-1}$, (b) stopped-flow conditions (with a pre-excitation flow rate of $v = 2$ cm/s), and (c) non-zero flow- and reaction-rates post-excitation ($v = 2$ cm/s and $k = 1$ s$^{-1}$). The spatially-varying signal intensity is reflective of the zero-order kinetics occurring pre- and post-excitation. A total of 256 gradient echoes were included, each composed of 128 points and lasting 0.5 ms. The image field-of-view was set to 2.3 cm and the gradient amplitude equalled 26.14 G/cm. The chemical shift of the reactant and product were simulated to be $\omega_A/2\pi = -40$ Hz and $\omega_B/2\pi = +50$ Hz, respectively.